\documentclass{emulateapj}

\usepackage{natbib}
\usepackage{amsmath}

\slugcomment{Accepted for publication in \textit{The Astrophysical Journal}}

\shorttitle{Morphologies and CMR in J$2215.9-1738$ at $z=1.46$}
\shortauthors{Hilton et al.}

\begin{document}

\title{The \textit{XMM} Cluster Survey: Galaxy Morphologies and the \\Color--Magnitude Relation in XMMXCS J$2215.9-1738$ at \lowercase{$z=1.46$}}

\author{Matt Hilton\altaffilmark{1, 2, 3}, S. Adam Stanford\altaffilmark{4, 5}, John P. Stott\altaffilmark{3}, Chris A. Collins\altaffilmark{3}, 
Ben Hoyle\altaffilmark{6}\\
Michael Davidson\altaffilmark{7}, Mark Hosmer\altaffilmark{8}, Scott T. Kay\altaffilmark{9}, Andrew R. Liddle\altaffilmark{8}, Ed Lloyd-Davies\altaffilmark{8}\\
Robert G. Mann\altaffilmark{7}, Nicola Mehrtens\altaffilmark{8}, Christopher J. Miller\altaffilmark{10}, Robert C. Nichol\altaffilmark{6}\\
A. Kathy Romer\altaffilmark{8}, Kivanc Sabirli\altaffilmark{8}, Martin Sahl{\'e}n\altaffilmark{8}, Pedro T. P. Viana\altaffilmark{11,12}\\
Michael J. West\altaffilmark{13}, Kyle Barbary\altaffilmark{14,15}, Kyle S. Dawson\altaffilmark{16}, Joshua Meyers\altaffilmark{14,15}\\
Saul Perlmutter\altaffilmark{14,15}, David Rubin\altaffilmark{14,15}, Nao Suzuki\altaffilmark{14}\\
}

\altaffiltext{1}{Astrophysics \& Cosmology Research Unit, School of Mathematical Sciences, University of 
KwaZulu-Natal, Private Bag X54001, Durban 4000, South Africa; hiltonm@ukzn.ac.za}
\altaffiltext{2}{South African Astronomical Observatory, PO Box 9, Observatory, 7935, Cape Town, South Africa}
\altaffiltext{3}{Astrophysics Research Institute, Liverpool John Moores University, Twelve Quays House, Egerton Wharf, Birkenhead, CH41 1LD, UK}
\altaffiltext{4}{University of California, Davis, CA 95616, USA}
\altaffiltext{5}{Institute of Geophysics and Planetary Physics, Lawrence Livermore National Laboratory,
Livermore, CA 94551, USA}
\altaffiltext{6}{Institute of Cosmology and Gravitation, University of Portsmouth, Portsmouth, PO1 2EG, UK}
\altaffiltext{7}{Institute of Astronomy, University of Edinburgh, Blackford Hill, Edinburgh, EH9 9HJ, UK}
\altaffiltext{8}{Astronomy Centre, University of Sussex, Falmer, Brighton, BN1 9QH, UK}
\altaffiltext{9}{University of Manchester, Jodrell Bank Observatory, Macclesfield, Cheshire, SK11 9DL, UK}
\altaffiltext{10}{Cerro-Tololo Inter-American Observatory, National Optical Astronomy Observatory, 950 North
Cherry Ave., Tucson, AZ 85719, USA}
\altaffiltext{11}{Departmento de Matem\'{a}tica Aplicada da Faculdade de Ci\^{e}ncias da Universidade do Porto, Rua do Campo Alegre, 687, 4169-007, Portugal}
\altaffiltext{12}{Centro de Astrof\'{\i}sica da Universidade do Porto, Rua das Estrelas, 4150-762, Porto, Portugal}
\altaffiltext{13}{European Southern Observatory, Alonso de C{\'o}rdova 3107, Vitacura, Casilla 19001, Santiago 19, Chile}
\altaffiltext{14}{E. O. Lawrence Berkeley National Laboratory, 1 Cyclotron Road, Berkeley, CA 94720, USA}
\altaffiltext{15}{Department of Physics, University of California Berkeley, Berkeley, CA 94720-7300, USA}
\altaffiltext{16}{Department of Physics and Astronomy, University of Utah, Salt Lake City, Utah 84112, USA}

\begin{abstract}
We present a study of the morphological fractions and color--magnitude relation in the most distant X-ray selected galaxy cluster currently known, XMMXCS J$2215.9-1738$ at $z=1.46$, using a combination of optical imaging data obtained with the \textit{Hubble Space Telescope} Advanced Camera for Surveys, and infrared data from the Multi-Object Infrared Camera and Spectrograph, mounted on the 8.2m \textit{Subaru} telescope. We find that the morphological mix of the cluster galaxy population is similar to clusters at $z \sim 1$. Within the central 0.5 Mpc, approximately $\sim 62$\% of the galaxies identified as likely cluster members are ellipticals or S0s; and $\sim 38$\% are spirals or irregulars. Therefore early type galaxies were already entrenched as the dominant galaxy population in at least some clusters approximately $\sim 4.5$ Gyr after the Big Bang. We measure the color--magnitude relations for the early type galaxies, finding that the slope in the $z_{850}-J$ relation is consistent with that measured in the Coma cluster, some $\sim 9$ Gyr earlier, although the uncertainty is large. In contrast, the measured intrinsic scatter about the color--magnitude relation is more than three times the value measured in Coma, after conversion to rest frame $U-V$. From comparison with stellar population synthesis models, the intrinsic scatter measurements imply mean luminosity weighted ages for the early type galaxies in J$2215.9-1738$ of $\approx 3$ Gyr, corresponding to the major epoch of star formation coming to an end at $z_f \approx 3-5$. We find that the cluster exhibits evidence of the `downsizing' phenomenon: the fraction of faint cluster members on the red sequence expressed using the Dwarf-to-Giant Ratio (DGR) is $0.32\pm 0.18$ within a radius of $0.5R_{200}$. This is consistent with extrapolation of the redshift evolution of the DGR seen in cluster samples at $z < 1$. In contrast to observations of some other $z > 1$ clusters, we find a lack of very bright galaxies within the cluster.
\end{abstract}


\section{Introduction}
\label{s_intro}
The galaxy populations of clusters are dominated by galaxies of early morphological type, ellipticals and S0s, which form a tight color--magnitude relation (CMR) or `red sequence'. For decades, this relation was interpreted as evidence that the stars in early type galaxies in clusters are uniformly old, being formed at redshift $z_f > 2$, with the slope of the relation being primarily due to a mass-metallicity relation \citep{Larson_1974, Tinsley_1978, Gallazzi_2006}. The alternative interpretation of the CMR as being predominantly an age sequence was conclusively ruled out by observations of clusters at $z \lesssim 0.4$, which showed that the slope of the CMR evolves little with redshift \citep[][]{KodamaArimoto_1997}.

A number of studies of clusters conducted with the \textit{Hubble Space Telescope} have shown that the CMR of elliptical galaxies remains well established at progressively higher redshifts \citep[e.g.][]{Ellis_1997, Stanford_1998, vanDokkum_2000}, at least up to $z\sim1.3$ \citep{vanDokkumLynx_2001, Blakeslee_2003, MeiLynx_2006}. The only study to date of the CMR in a cluster at $z=1.4$ is consistent with this picture \citep{Lidman_2008}. Measurements of the intrinsic scatter about the CMR can be used to constrain the ages of the constituent stellar populations of cluster early type galaxies, and indicate that major epoch of star formation in clusters was completed at $z > 2$ \citep[e.g.][]{Bower_1992, Blakeslee_2003, Mei_2009}. Studies of the fundamental plane of elliptical galaxies up to high redshift similarly indicate that the bulk of stellar populations in elliptical galaxies were formed at $z > 2$ \citep[e.g.][]{vanDokkumStanford_2003, Holden_2005, Jorgensen_2006, vanDokkumVanDerMarel_2007}.

These observations are consistent with a simple formation scenario for elliptical galaxies, in which they formed the bulk of their stellar mass in a single event at high redshift, and evolved passively thereafter. However, the latest semi-analytic models of galaxy formation, constructed within the hierarchical formation paradigm, are also able to successfully reproduce the old ages of stellar populations in elliptical galaxies \citep{DeLucia_2006, Menci_2008}.

Recent observations over a wide range in redshift in both clusters and the field have revealed that star formation appears to be completed earlier in higher mass galaxies than in low mass galaxies, a phenomenon dubbed `downsizing' \citep{Cowie_1996}. In color-magnitude diagrams of high redshift clusters, this effect manifests itself as a deficit of faint galaxies on the red sequence in comparison to clusters observed at lower redshift \citep[e.g.][]{DeLucia_2004, DeLuciaCMR_2007, Stott_2007}, and may have an environmental dependence in the sense that the faint end of the CMR is populated at earlier times in denser regions, such as in the cluster cores, compared to low density regions such as groups \citep[e.g.][]{Tanaka_2005, Tanaka_2008, Gilbank_2008}.

This could be explained qualitatively by a scenario in which the faint end of the red sequence is being built up by the transformation of star forming, spiral galaxies into passively evolving S0s as redshift increases. Up to $z\sim 1$, studies of magnitude-limited samples of cluster galaxies have revealed that the elliptical galaxy fraction within clusters is found to remain roughly constant, while the fraction of S0 galaxies decreases, with a corresponding increase in the fraction of spiral and irregular galaxies \citep{Dressler_1997, Smith_2005, Postman_2005}. When considering stellar mass-limited samples, it has been found that the fraction of massive, early type galaxies as a whole remains roughly constant, at least up to $z\approx 0.8$ \citep{Holden_2006, Holden_2007, vanDerWel_2007}. This suggests that the much stronger evolution seen in magnitude-limited samples is dominated by fainter, lower mass (sub-$M^*$ in the galaxy stellar mass function) galaxies. It should be noted that at the current time the evidence for downsizing within the centers of rich clusters is still somewhat contentious (\citealt{Andreon_2008}; \citealt*{Crawford_2009}).

In this paper, we present a study of the galaxy morphologies and color--magnitude relation in XMMXCS J$2215.9-1738$ at $z=1.46$ \citep{Stanford_2006, Hilton_2007}, the most distant X-ray selected galaxy cluster currently known. To date, only five other spectroscopically confirmed clusters are known at $z > 1.3$, XMMU J$2235.3-2557$ at $z=1.39$, discovered serendipitously with \textit{XMM-Newton} \citep{Mullis_2005}; three clusters at $z=1.33-1.41$ from the IRAC Shallow Cluster Survey \citep[ISCS;][]{Eisenhardt_2008, Stanford_2005}; and the $z=1.335$ cluster J$003550-431224$, discovered by the \textit{Spitzer} Adaptation of the Red-sequence Cluster Survey \citep[SpARCS;][]{Wilson_2008}.

J$2215.9-1738$ was discovered as part of the ongoing optical follow up campaign to the \textit{XMM} Cluster Survey \citep[XCS\footnotemark;][]{Romer_2001}, which has the primary aim of constraining the cosmological parameters through measuring the evolution of the cluster mass function with redshift. Predictions for the expected cosmological constraints that are expected to be achieved by the full survey can be found in \citet{Sahlen_2008}. J$2215.9-1738$ has X-ray luminosity $L_{\rm X} = 4.39^{+0.46}_{-0.37} \times 10^{44}$ ergs s$^{-1}$, temperature $T = 7.4^{+1.6}_{-1.1}$ keV \citep{Stanford_2006}, and velocity dispersion $\sigma_v = 580 \pm 140$ km s$^{-1}$ \citep{Hilton_2007}. There is mild evidence that the cluster velocity distribution is bimodal, which, if confirmed by further spectroscopic observations, would indicate that the cluster is undergoing a merger close to the line of sight \citep{Hilton_2007}.\footnotetext{http://www.xcs-home.org}

The structure of this paper is as follows. We begin by describing the observations and data reduction in \textsection~\ref{s_observationsDataReduction}. In \textsection~\ref{s_analysis}, we describe our photometric measurements, morphological classification, and photometric redshift selection of the cluster members. We present the morphological fractions, fits to the color--magnitude relation, and inferred ages for the stellar populations of the early type galaxies in \textsection~\ref{s_results}. Finally, we discuss our findings in comparison with previous work in \textsection~\ref{s_discussion}.

We assume a concordance cosmology of $\Omega_m=0.3$, $\Omega_\Lambda=0.7$, and $H_0=70$ km s$^{-1}$ Mpc$^{-1}$ throughout, where $\Omega_\Lambda$ is the energy density associated with a cosmological constant.

\section{Observations and data reduction}

\label{s_observationsDataReduction}
\begin{figure}
\epsscale{0.9}
\plotone{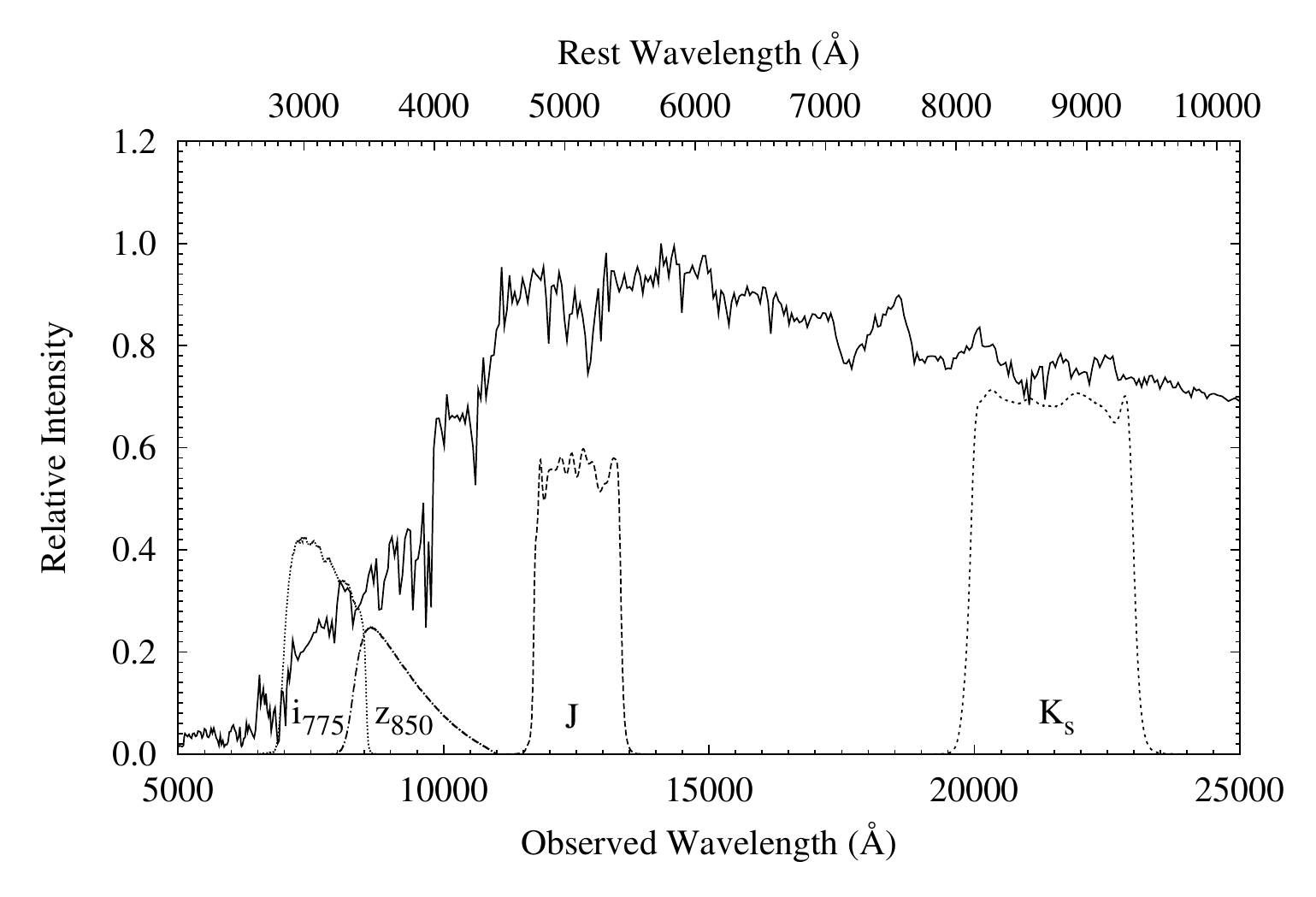}
\caption{Response curves for the ACS ($i_{775}$, $z_{850}$) and MOIRCS ($J$, $K_s$) filters used in this study. The solid line is a \citet{BruzualCharlot_2003} model spectrum of a galaxy with solar metallicity, \citet{Salpeter_1955} Initial Mass Function, formed at redshift $z_f=4.5$ in a 0.1 Gyr burst of star formation, as it would be observed at the cluster redshift of $z=1.46$. The near infrared data are essential to bracket the 4000 \AA{} break in early type galaxies at this redshift.}
\label{f_passbandsPlot}
\end{figure}

For our study of the galaxy populations of J2215.9$-$1738, we use a combination of optical data in the $i_{775}$ and $z_{850}$ bands obtained with the \textit{Hubble Space Telescope} (\textit{HST}) Advanced Camera for Surveys \citep[ACS;][]{Ford_2003}, and near infrared data in the $J$ and $K_s$ bands taken using the Multi-Object Infrared Camera and Spectrograph \citep[MOIRCS;][]{Ichikawa_2006} on the 8.2 m \textit{Subaru} telescope at the summit of Mauna Kea, Hawaii. As shown in Fig.~\ref{f_passbandsPlot}, at the cluster redshift of $z=1.46$ the optical data lies blueward of the 4000 \AA{} break, and therefore near infrared data are essential to provide a color which is able to separate the cluster red sequence from bluer, star forming galaxies. The $z_{850}-J$ color at the cluster redshift of $z=1.46$ roughly corresponds to rest frame $U-V$, while $z_{850}-K_s$ is roughly equivalent to rest frame $U-z$.

\begin{figure}
\epsscale{0.9}
\plotone{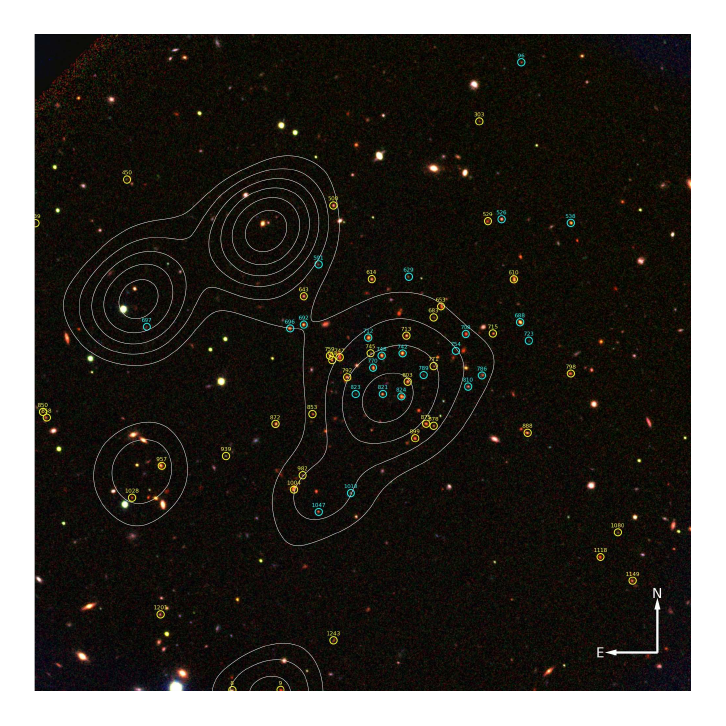}
\caption{Color-composite ($z_{850}$, $J$, $K_s$) using the ACS and MOIRCS imaging data presented in this paper, centered on the cluster X-ray coordinates of $22^{h}15^{m}58.5^{s}$, $-17^\circ38^\prime02.5^{\prime\prime}$ (J2000). The image is 3.04$^{\prime}$ on a side. The ACS $z_{850}$ image has been degraded to match the 0.5$^{\prime\prime}$ resolution of the MOIRCS $J$, $K_s$ imaging. X-ray contours are overlaid in white; cyan points indicate spectroscopically confirmed cluster members; yellow points indicate additional cluster members selected using photometric redshifts.}
\label{f_RGBImage}
\end{figure}

Fig.~\ref{f_RGBImage} presents a color-composite ($z_{850}$, $J$, $K_s$) image of the cluster. It is apparent from Fig.~\ref{f_RGBImage} that J$2215.9-1738$ lacks an obvious Brightest Cluster Galaxy (BCG), in contrast with the only other similarly high redshift X-ray selected cluster currently known, XMMU J$2235.3-2557$ at $z=1.39$ \citep[][see also \citealt{Collins_2009}]{Lidman_2008}. Similarly, the galaxy distribution is much less compact in J$2215.9-1738$ than in J$2235.3-2557$.

We now describe the observations and data reduction performed for each instrument in turn.

\subsection{Hubble Space Telescope}
J$2215.9-1738$ was observed using the ACS Wide Field Channel (WFC) as part of a program designed to place constraints on the dark energy through observations of high redshift Type Ia supernovae\footnotemark. \footnotetext{Based on observations made with the NASA/ESA Hubble Space Telescope, obtained from the data archive at the Space Telescope Institute. STScI is operated by the association of Universities for Research in Astronomy, Inc. under the NASA contract NAS 5-26555. The observations are associated with program 10496.} The field of view of the ACS WFC is approximately $3.4^\prime \times 3.4^\prime$, and the pixel scale of the detectors is $0.05^{\prime\prime}$ per pixel. A detailed description of the ACS observations is presented in \citet{Dawson_2008}. A total of 3320 sec of exposure was obtained in $i_{775}$ across 12 frames. In $z_{850}$, 16935 sec of exposure was obtained in 64 frames. In both cases, the final stacked images were created using the \texttt{MULTIDRIZZLE} \texttt{PyRAF} task\footnotemark. \footnotetext{http://www.stsci.edu/resources/software\_hardware/ ...\\
... pyraf/stsci\_python}

\subsection{Subaru}
J2215.9$-$1738 was observed with MOIRCS in photometric conditions on UT 2007 August 08. The observations were carried out in service mode. The field of view of MOIRCS is $4^\prime \times 7^\prime$, which is imaged at a resolution of $0.117^{\prime\prime}$ per pixel on to a pair of $2048 \times 2048$ HAWAII-2 detector arrays. Unfortunately, one of the detector arrays was inoperable at the time our observations were carried out. Our observations were designed to place the target cluster at the center of the second, working detector array. We performed imaging in the $J$ and $K_s$ bands, using a 9-point circular dither pattern of radius 25$^{\prime\prime}$ to ensure good sky subtraction. We obtained a total of 1485 sec of integration in $J$, for an individual frame time of 165 seconds at each dither position. In the $K_s$ band, the total integration time was 1242 sec, achieved by co-adding $3\times 46$ sec exposures at each dither position. The observations were obtained at airmass $\approx 1.3$. The seeing was excellent, at $\approx 0.5^{\prime\prime}$ in both $J$ and $K_s$.

The data were reduced using the \texttt{MCSRED} package for the \texttt{IRAF}\footnotemark environment in the standard manner: the individual dither frames were flat fielded, sky subtracted, corrected for distortion caused by the optical design of MOIRCS, and registered to a common pixel coordinate system. The final science images were created from the $3\sigma$-clipped mean of the dither frames.\footnotetext{IRAF is distributed by the National Optical Astronomy Observatories, which are operated by the Association of Universities for Research in Astronomy, Inc., under cooperative agreement with the National Science Foundation.}

\section{Analysis}
\label{s_analysis}

\subsection{Photometry}
\label{s_analysisPhotometry}
As most of our ACS data were obtained before UT 2006 July 04, we adopted AB photometric zeropoints in the $i_{775}$ and $z_{850}$ ACS bands of 25.678 and 24.867 respectively\footnotemark. After this date, the ACS photometric zeropoints changed due to a change in the temperature of CCDs: frames obtained after this change occurred were adjusted to match the above zeropoints. During the MOIRCS observations, we obtained imaging of several photometric standards from the UKIRT faint standards list. The zeropoints we determined in the $J$ and $K_s$ bands are 26.435$\pm$0.007 and 25.900$\pm$0.010 on the Vega system respectively. For the near infrared data we adopt conversions to the AB magnitude system \citep{Oke_1974} of $J (\rm AB) = J (\rm Vega)+0.943$ and $K_s (\rm AB) = K_s (\rm Vega) + 1.86$ \citep{TokunagaVacca_2005}. All magnitudes quoted throughout the rest of this paper are on the AB system, unless stated otherwise.\footnotetext{see \texttt{http://www.stsci.edu/hst/acs/analysis/zeropoints}}

To measure accurate colors for the galaxies in J2215.9$-$1738, we need to perform measurements through a consistent set of photometric apertures. We used the \texttt{SEXTRACTOR} package \citep{BertinArnouts_1996} in dual image mode to achieve this, where objects are detected and apertures are defined using the first image, while photometry is performed on the second image. We chose to use the MOIRCS $K_s$ image as the detection image. As the ACS and MOIRCS imaging data have different angular and pixel resolution, some additional processing was required prior to performing the photometry.

Firstly, we trimmed the ACS and MOIRCS images to cover a common area of dimensions $3.04^\prime \times 3.04^\prime$, centered on the cluster X-ray position. This area encloses the overlapping ACS coverage from the individual dither frames, rejecting higher noise regions at the edges covered by fewer ACS pointings. We then scaled the MOIRCS images from their native pixel scale of 0.117$^{\prime\prime}$ per pixel to the ACS pixel scale of 0.05$^{\prime\prime}$ per pixel, interpolating using a 3rd order spline polynomial. The MOIRCS images were then registered to the ACS images to pixel accuracy. Finally, the ACS images were matched to the 0.5$^{\prime\prime}$ angular resolution of the MOIRCS images by convolution with a Gaussian filter with $\sigma = 4.0$ pixels.

To estimate the photometric errors in each band, we constructed RMS images for use as weight images by \texttt{SEXTRACTOR}. For the MOIRCS images, we used the RMS images produced by the \texttt{MCSRED} data reduction package - these are simply the standard deviation at each pixel derived from the 9 individual dither frames. For the ACS images, the RMS images were constructed from the drizzled science images and the exposure time maps output by the \texttt{MULTIDRIZZLE} task, taking into account the photon statistics and the detector read noise. The drizzling process used to make the final ACS science images correlates the pixel-to-pixel noise, which, if ignored, would lead to underestimated photometric errors. We corrected for this effect following the methodology outlined in the Appendix of \citet{Casertano_2000}.

The effect of the interpolation operations on the photometry from the MOIRCS images was tested by cross matching catalogs constructed at the native MOIRCS pixel scale, with catalogs made from the images resampled to match the ACS pixel scale. Objects were matched between each catalog within a 0.5$^{\prime\prime}$ radius and the $1.0^{\prime\prime}$ aperture magnitudes were compared. We found that the median magnitude difference between the two catalogs was $\approx 0.001$ for $J, K_s < 24$. We decided to treat the scatter between the catalogs as an additional source of uncertainty: we fitted an exponential model to the scatter as a function of magnitude, applied this model to the $J, K_s$ aperture magnitudes in the catalog constructed at the ACS pixel resolution, and added the results in quadrature to the photometric errors output by \texttt{SEXTRACTOR}. The increase in the $J$ and $K_s$ aperture magnitude uncertainties resulting from this procedure is small, $ < 0.04$ mag. at $ J, K_s < 24$.

To classify objects as stars and galaxies, we constructed a second set of catalogs in the manner described above, with the exception that the ACS images were not matched to the 0.5$^{\prime\prime}$ resolution of the MOIRCS images, and instead were read into \texttt{SEXTRACTOR} at their native $\approx 0.09^{\prime\prime}$ angular resolution. We used the neural-network based star-galaxy classification provided by \texttt{SEXTRACTOR} (\texttt{CLASS\_STAR}) to determine the object types. We classify all objects with \texttt{CLASS\_STAR}$<0.9$ as measured in the $z_{850}$ image as galaxies; all other objects are assumed to be stars. We chose to carry out the classification in the $z_{850}$ band, because the $z_{850}$ image is significantly deeper than the $i_{775}$ image, reaching an approximate $5\sigma$ limiting magnitude of $\approx 26.0$ for galaxies, compared to $\approx 25.1$ for the $i_{775}$ image (estimated from the photometric uncertainties).

We corrected the photometry for the effect of Galactic extinction using the dust maps and software of \citet*{Schlegel_1998}. In all that follows, \texttt{SEXTRACTOR} \texttt{MAG\_AUTO} magnitudes were adopted as measurements of total galaxy magnitudes, and colors were measured in 1.0$^{\prime\prime}$ diameter circular apertures, equal to twice the seeing of the MOIRCS images. This is comparable to the typical diameter of the cluster member galaxies.

Note that we do not attempt to correct the photometry of objects blended as a consequence of performing the object detection in the $K_s$-band. Only a small number of such objects are seen, perhaps due to the galaxy distribution in J$2215.9-1738$ being less compact than other similarly high redshift clusters, as noted in \textsection~\ref{s_observationsDataReduction} above. As shown in \textsection~\ref{s_resultsMorphologies}, this has a negligible effect on our results.

\subsection{Morphologies}
\label{s_analysisMorphologies}
Morphological classification was performed on all 201 galaxies with magnitudes brighter than $z_{850} \leq 24$, by visual inspection of the $z_{850}$-band imaging data. This is $\sim2$ magnitudes brighter than the approximate $5\sigma$ magnitude limit of the $z_{850}$-band image. Due to the high redshift of J$2215.9-1738$, the morphological classification was carried out in the rest frame $U$-band, rather than the traditional $B$-band \citep[e.g.][]{Postman_2005}. Furthermore, the typical diameter of the cluster member galaxies in the ACS imaging is $< 20$ pixels, making detailed morphological classification on the traditional Hubble system difficult. For these reasons, we chose to place the galaxies into four broad morphological bins: elliptical (E; corresponding to E1$-$E9 on the traditional Hubble system), lenticular (S0), spiral (Sp; Sa$-$Sd, SBa$-$SBd on the traditional Hubble system) or irregular (Irr).

The classification was carried out on the complete sample by a team of five classifiers (CAC, MH, BH, SAS, JPS). Each galaxy in the sample was examined by means of a 150 pixel on a side postage stamp image, centered on the galaxy in question. No further information on the galaxy to be classified (e.g., position, redshift, color or magnitude) was provided during the classification process, in order to ensure that the classification was determined strictly on the evidence of the imaging data alone, guarding against potential biases. The same training set of galaxy images, constructed from the \citet{Postman_2005} study of galaxy morphologies in $0.8 < z < 1.3$ clusters, was provided to each classifier.

We found there was generally good agreement on the morphological classification of the individual galaxies among the five classifiers: majority agreement (i.e. $> 50$\% of the votes cast by the classifiers were for a particular morphology) was reached for 85\% of the sample. Agreement by $> 2/3$ of the classifiers was reached for 67\% of the sample. We estimated the RMS scatter between the morphological fractions determined by each classifier and adopted this as our estimate of the morphological classification error, used in \textsection~\ref{s_resultsMorphologies} below. For the elliptical galaxy fraction, we measure a scatter of $\sigma_{f_E} = 0.11$; for the S0 galaxy fraction, we obtain $\sigma_{f_{S0}} = 0.05$; and for the late type galaxy fraction (Sp+Irr), we obtain $\sigma_{f_{Sp+Irr}} = 0.11$. The uncertainty in the early type galaxy fraction (E+S0) is $\sigma_{f_{E+S0}} = 0.11$. These uncertainties are comparable in size to those obtained by \citet{Postman_2005}.

Note that in this study we have not taken into account the potential impact of surface brightness dimming, the effect of which can impair the ability to distinguish between, e.g. ellipticals/S0s, and S0/Sa galaxies. \citet{Postman_2005} conducted simulations in order to quantify the size of this effect in their ACS study of the morphology--density relation in $z\sim 1$ clusters, and found that the effect of surface brightness dimming was successfully mitigated by their chosen exposure times (typically $\approx 12000$ sec in $z_{850}$). We therefore expect that the impact of surface brightness dimming on our results is alleviated to some extent by our long exposure time, $\approx 40$\% longer than the typical exposure time used by \citealt{Postman_2005}, and by the broad morphological bins we have adopted.

\subsection{Photometric redshifts}
\label{s_photoZs}
Secure spectroscopic redshifts have been obtained for a total of 58 galaxies within the $3.04^\prime$ field covered by the ACS and MOIRCS observations. The majority of these redshifts come from observations of J$2215.9-1738$ using the DEep Imaging Multi-Object Spectrograph \citep[DEIMOS;][]{Faber_2003} on the 10m Keck II telescope, and the Focal Reducer and Low Dispersion Spectrograph \citep[FORS2;][]{Appenzeller_1998} on the 8m Very Large Telescope (VLT), and are described in \citet{Hilton_2007}. Observations through a further three DEIMOS slit masks have been obtained subsequent to this work; one being observed in September 2007, and two in September 2008. Details of the 2007 and 2008 Keck observations will be presented in S. A. Stanford et al.~(2009, in preparation). Of the 58 galaxies with spectroscopic redshifts, 24 are cluster members, and so we chose to use the available four band ($i_{775}$, $z_{850}$, $J$, $K_s$) photometry to determine cluster membership using photometric redshifts, for all galaxies in the $K_s$-selected catalog described in \textsection~\ref{s_analysisPhotometry} above.

We used the spectral template fitting code \texttt{EAZY} \citep*{Brammer_2008} to estimate photometric redshifts. \texttt{EAZY} has been designed to be especially useful in cases where spectroscopic redshifts are only available for a biased subset of galaxies. This is the case for our dataset, as color, magnitude criteria were used to preferentially select likely cluster members as spectroscopic targets. Furthermore, the limited size of our spectroscopic subsample makes the use of `training set'-type techniques inappropriate, such as, e.g. neural-network based redshift estimates \citep{Collister_2004}, or iterative correction of empirical spectral templates \citep{Feldmann_2006}. We used the default set of five spectral templates included with \texttt{EAZY}, which is constructed by applying the non-negative matrix factorization technique of \citet{Blanton_2007} to a set of 3,000 stellar population models from the PEGASE library \citep{Fioc_1997}, and a catalog of synthetic galaxy photometry derived from the semi-analytic simulation of \citet{DeLucia_2007}. We chose to use the option within \texttt{EAZY} to fit a linear combination of all of the templates to our galaxy catalog, over a redshift range of $0 < z < 4$, using the $1.0^{\prime\prime}$ aperture magnitudes as the input photometry. Finally, we chose to use the $K-$magnitude based Bayesian redshift prior included with \texttt{EAZY}, which is derived from the synthetic, semi-analytic model galaxy photometry used to define the template set, and adopted the maximum likelihood redshift estimate after the application of this prior as the redshift estimate $z_p$ for each galaxy.

\begin{figure}
\epsscale{0.75}
\plotone{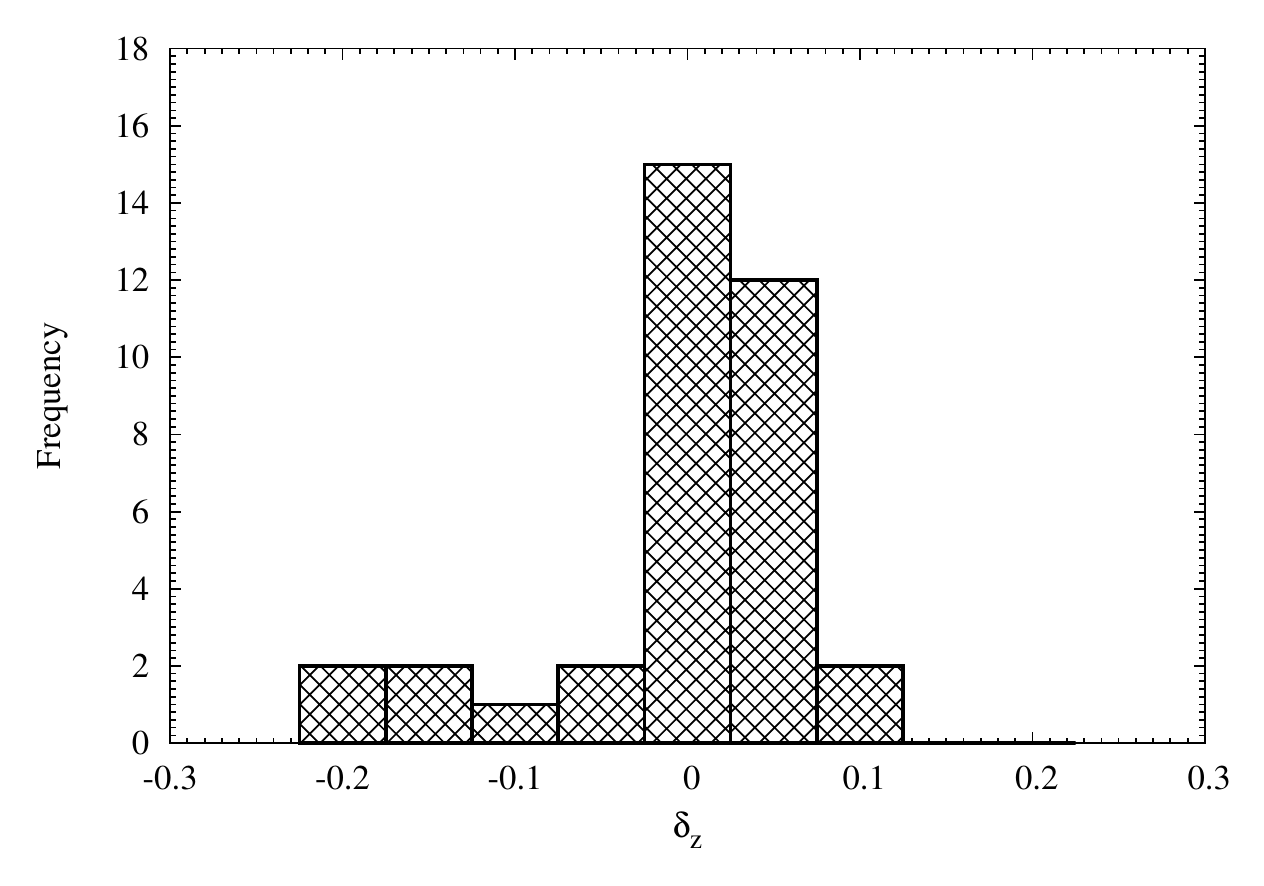}
\caption{Histogram of the photometric redshift residuals $\delta z = (z_s - z_p) / (1 + z_s)$, where $z_s$ is the spectroscopic redshift, and $z_p$ is the corresponding photometric redshift. All galaxies with spectroscopic redshifts $z_s > 1$ are included. The photometric redshifts are almost unbiased; the median of the distribution is $\delta z = +0.015$.}
\label{f_zpzsHistogram}
\end{figure}

We checked the accuracy of the photometric redshift estimates produced by \texttt{EAZY} by comparison with the spectroscopic subsample. Given the wavelength coverage of our available photometric bands, we would not expect to obtain reliable photometric redshifts for galaxies with redshifts below $z < 1$, as photometric redshift techniques rely on identifying strong spectral breaks in order to assign redshift estimates, and the $i_{775}$-band only drops below 4000 \AA{} in the rest frame at $z \gtrsim 1$. We therefore performed the check on the photometric redshift accuracy on the 36 galaxies with spectroscopic redshifts $z_s > 1$. We found that the typical scatter $\sigma_{\delta_z}$ in the photometric redshift residuals $\delta z = (z_s - z_p) / (1 + z_s)$ was $\sigma_{\delta_z} = 0.039$, where $\sigma_{\delta_z}$ was estimated using a biweight scale estimator \citep[e.g.][]{Beers_1990}, which is robust to outliers. The photometric redshifts appear to be almost unbiased in this redshift range, as the median $\delta z = +0.015$; however, a small number of objects have overestimated photometric redshifts, as can be seen from the negative tail of the distribution shown in Fig.~\ref{f_zpzsHistogram}.

A cut of $\mid z_p - z_c \mid < 2 \times \sigma_{\delta z} (1+z_c)$ was used to define cluster membership, where $z_c$ is the cluster redshift of $z=1.46$ - i.e. galaxies with $1.27 < z_p < 1.65$ were considered to be members of the cluster. We used the spectroscopic subsample to assess the amount of contamination from galaxies with spurious photometric redshifts (i.e. with $z_s < 1.27$, $z_s > 1.65$); contamination from interlopers, defined as galaxies with velocities outside of the range $\pm 3 \sigma_v$ around the cluster redshift, where $\sigma_v$ is the line of sight velocity dispersion measured in \citet{Hilton_2007} (i.e. $z_s > 1.46+0.015$ or $z_s < 1.46-0.015$, but $1.27 < z_s < 1.65$); and missing spectroscopically identified members (i.e. with $z_p < 1.27$, $z_p > 1.65$ and $1.445 < z_s < 1.475$).

Initially, using only the cut in $z_p$ to define membership, a total of 13 galaxies (22\% of the spectroscopic subsample) were identified as having spurious photometric redshifts, all with $z_s < 1$. However, galaxies with unreliable photometric redshift estimates have flat or multi-modal posterior redshift probability distributions, and can be easily identified by their $p_{\Delta_z}$ value \citep[introduced by][and labelled \texttt{ODDS} in the output from \texttt{EAZY}]{Benitez_2000}, which is the fraction of the integrated probability that lies within $\pm \Delta z$ of the photometric redshift estimate $z_p$ (in the case of \texttt{EAZY}, $\Delta_z=0.2$). Galaxies with lower $p_{\Delta_z}$ values have broader redshift probability distributions, and correspondingly less reliable photometric redshift estimates. Applying a cut of $p_{\Delta_z} > 0.9$ reduces the number of spurious photometric redshifts in the spectroscopic subsample to two galaxies (3.4\% of the spectroscopic subsample). We therefore chose to use this cut in $p_{\Delta_z}$ in all that follows. A more conservative cut of $p_{\Delta_z} > 0.95$ reduces the number of galaxies with spurious photometric redshifts to just a single galaxy, but this comes at the cost of a smaller sample size and increased error bars on the derived color--magnitude relation parameters (see \textsection~\ref{s_resultsCMRs}).

The number of spectroscopically confirmed non cluster members contaminating the photometrically selected sample was found to be three (10.3\% of the sample). However, four spectroscopically confirmed cluster members were missed by the photometric redshift selection (16.7\% of the sample of 24 confirmed cluster members). Using a broader cut in $z_p$ than we have applied naturally reduces the fraction of missed spectroscopic members, but at the cost of significantly increasing the contamination from spectroscopically confirmed interlopers. We show in \textsection~\ref{s_resultsMorphologies} and \textsection~\ref{s_resultsCMRs} that our measurements are robust to reasonable changes of the parameters governing the photometric redshift selection.

We further refined our photometric membership selection by excluding from the sample all galaxies with $K_s$-band magnitudes brighter than the BCG, i.e. $K_s < 20.57$. Although J$2215.9-1738$ lacks an obvious BCG, it possesses several candidates that are of similar brightness \citep[see][]{Collins_2009}, and this additional cut helps by removing seven bright, blue ($z_{850}-J < 1.0$), lower redshift galaxies from the sample. These objects are probably at $z\approx 1.3$, as indicated from their photometric redshifts, and one of these objects is a known interloper with spectroscopic redshift $z_s=1.301$. Finally, we removed from the sample all known interlopers with spectroscopic redshifts, and added back in the four spectroscopically confirmed members missed by the photometric redshift selection.

The final sample of photometrically selected members within the ACS/MOIRCS imaging area contains 64 galaxies. The colors, magnitudes, spectroscopic and photometric redshifts, and morphologies of the sample are recorded in Table~\ref{t_galaxyPhotometry}.

\section{Results}
\label{s_results}

\subsection{Morphologies}
\label{s_resultsMorphologies}

\begin{figure}
\epsscale{1.0}
\plotone{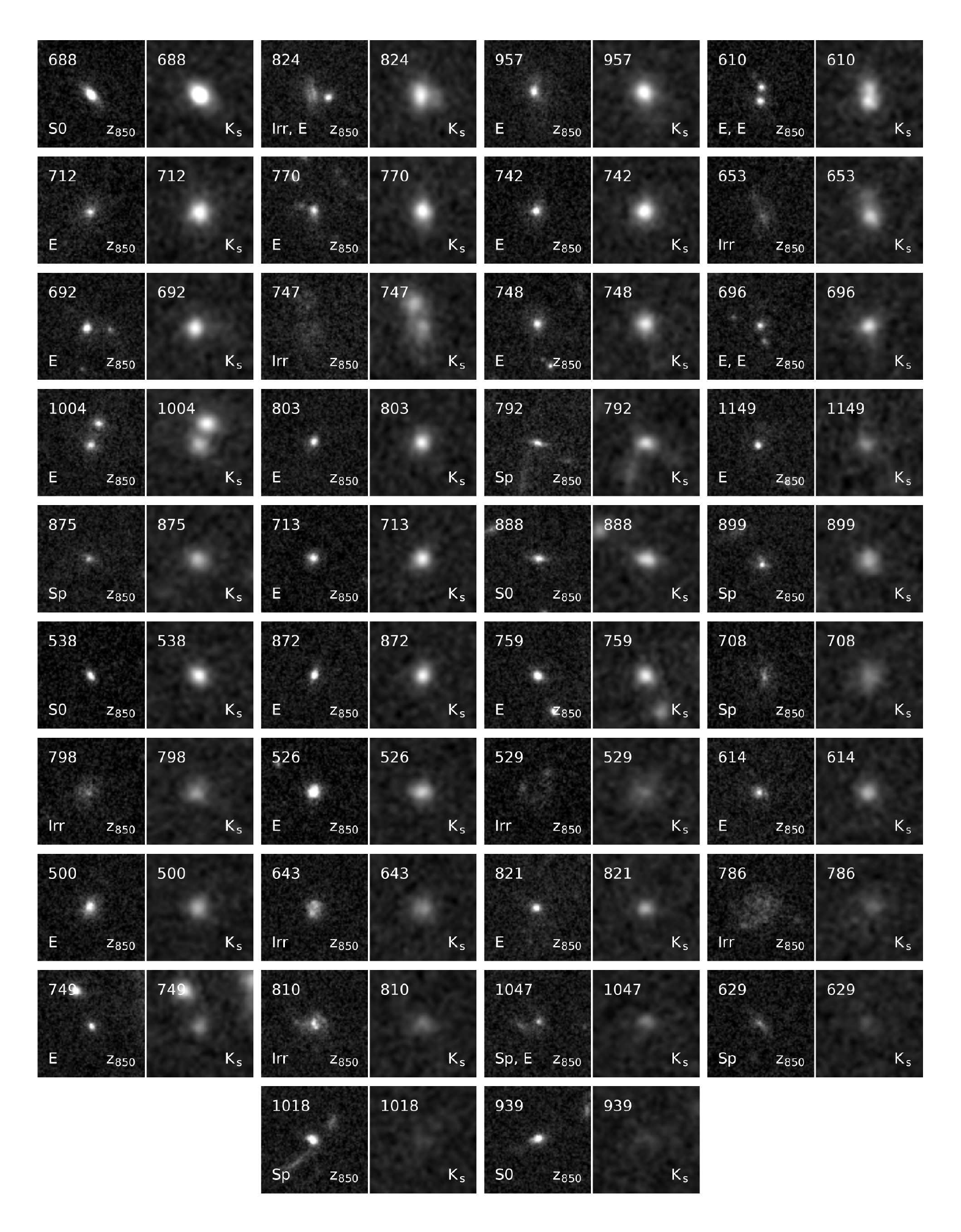}
\caption{Postage stamp images of all photometrically selected members of J$2215.9-1738$ for which morphological classification was carried out, located within a radius of 0.75 Mpc of the cluster X-ray position, and ordered by $K_s$ magnitude, from brightest (top left) to faintest (bottom right). The left postage stamp of each pair is taken from the ACS $z_{850}$-band image; the right hand image of each pair is taken from the MOIRCS $K_s$-band data which was used to perform the object detection (see \textsection~\ref{s_analysisPhotometry}). Each postage stamp is 3.75$^{\prime\prime}$ on a side, with East at the left. The number in the top left hand corner of each plot is the galaxy ID number in Table~\ref{t_galaxyPhotometry}; the galaxy morphology is indicated in the bottom left hand corner of each ACS $z_{850}$ image. Note that a few objects (IDs 824, 610, 696, and 1047) are blended in the $K_s$-selected catalog; these are assigned multiple morphologies, with the morphology of the easternmost component being quoted first.}
\label{f_morphologies}
\end{figure}

Fig.~\ref{f_morphologies} shows postage stamp images of all the galaxies with morphological classifications (i.e., all galaxies with $z_{850} \leq 24$), located within 0.75 Mpc of the cluster X-ray position, and selected as cluster members using the photometric criteria described in \textsection~\ref{s_photoZs} above. In addition to the $z_{850}$-band data from which the morphology of each object was determined, we also show a corresponding postage stamp taken from the $K_s$-band image that was used to perform the object detection (see \textsection~\ref{s_analysisPhotometry}). Only four objects in this sample were found to be blended in the $K_s$-selected catalog - these are IDs 824, 610, 696 and 1047, and are noted with a pair of morphologies in Fig.~\ref{f_morphologies} and Table.~\ref{t_galaxyPhotometry}.

Within a radius of 0.5 Mpc of the cluster X-ray position, there are a total of 39 galaxies with morphological classifications. The fraction of elliptical galaxies in this subsample is $f_E = 0.54 \pm 0.17$; the fraction of lenticular galaxies is $f_{S0} = 0.08 \pm 0.07$; and the fraction of late type galaxies is $f_{Sp+Irr} = 0.38 \pm 0.15$. The combined fraction of early type galaxies is $f_{E+S0} = 0.62 \pm 0.17$.

In calculating the errors on the morphological fractions quoted above, we add in quadrature the classification error (see \textsection~\ref{s_analysisMorphologies} above), the Poisson uncertainty, and the uncertainty in cluster membership due to the photometric redshift selection. To account for this latter effect, we perform 1000 Monte-Carlo simulations in which the photometric redshift of every galaxy is replaced by a random variate drawn from its photometric redshift probability distribution, as output by \texttt{EAZY}, and apply the algorithm described in \textsection~\ref{s_photoZs} to select cluster members. We find that the dominant source of error is the classification error; the photometric redshift selection increases the size of the uncertainties by at most $\approx$ 4\% for a given morphological bin.

To further check that the results are robust to the choice of parameters used to define the photometric redshift selection, we also calculated the morphological fractions for different photometric redshift cuts. For a more conservative choice of $p_{\Delta_z} > 0.95$, we obtain consistent results of $f_E = 0.56 \pm 0.17$; $f_{S0} = 0.06 \pm 0.07$; and $f_{Sp+Irr} = 0.38 \pm 0.16$, for a sample of 34 galaxies within a radius of 0.5 Mpc. Similarly, for a more stringent photometric redshift cut of $1.36 < z_p < 1.56$ with $p_{\Delta_z} > 0.9$, we obtain $f_E = 0.55 \pm 0.18$; $f_{S0} = 0.06 \pm 0.07$; and $f_{Sp+Irr} = 0.39 \pm 0.16$, for a sample of 31 galaxies within the same selection radius. Considering only the 19 galaxies with spectroscopic redshifts located within 0.5 Mpc of the cluster X-ray position, we obtain results consistent with those above, although with larger uncertainties. For this sample we find $f_E = 0.58 \pm 0.21$; $f_{S0} = 0.05 \pm 0.07$; and $f_{Sp+Irr} = 0.37 \pm 0.18$.

Given the large uncertainties, we are unable to examine the variation of the morphological fraction with radius. For example, extending the radius within which the morphological fractions are calculated to 0.75 Mpc does not change our results, as it only increases the sample size by a further three galaxies.

The object identified as the Brightest Cluster Galaxy (BCG) in J$2215.9-1738$, ID 688, has a clear S0 morphology as seen in Fig.~\ref{f_morphologies}. It is the brightest cluster member in the $K_s$-band (although there are in fact several other candidates of similar brightness), and has a spectrum consistent with a Luminous Red Galaxy (LRG) template \citep[see Fig.~1 of][]{Hilton_2007}. However, it lies $\approx 300$ kpc away from the cluster X-ray position. Although this would be unusual for clusters at low redshift, we note that the two brightest members of the $z=1.10$ cluster RDCS J$0910+5422$ also have S0 morphologies, and furthermore, are located at clustercentric distances $> 300$ kpc \citep{MeiRDCS_2006}. Therefore it is possible that J$2215.9-1738$ is not untypical of clusters at this epoch in this regard, given the small number of objects currently known at $z > 1$. A comparison of the stellar masses of BCGs in a sample of $z>1$ clusters, including J$2215.9-1738$, with the latest semi-analytic models of galaxy formation is presented in \citet{Collins_2009}.

\subsection{Color-magnitude diagrams}
\label{s_resultsCMDs}
\begin{figure}
\epsscale{0.61}
\plotone{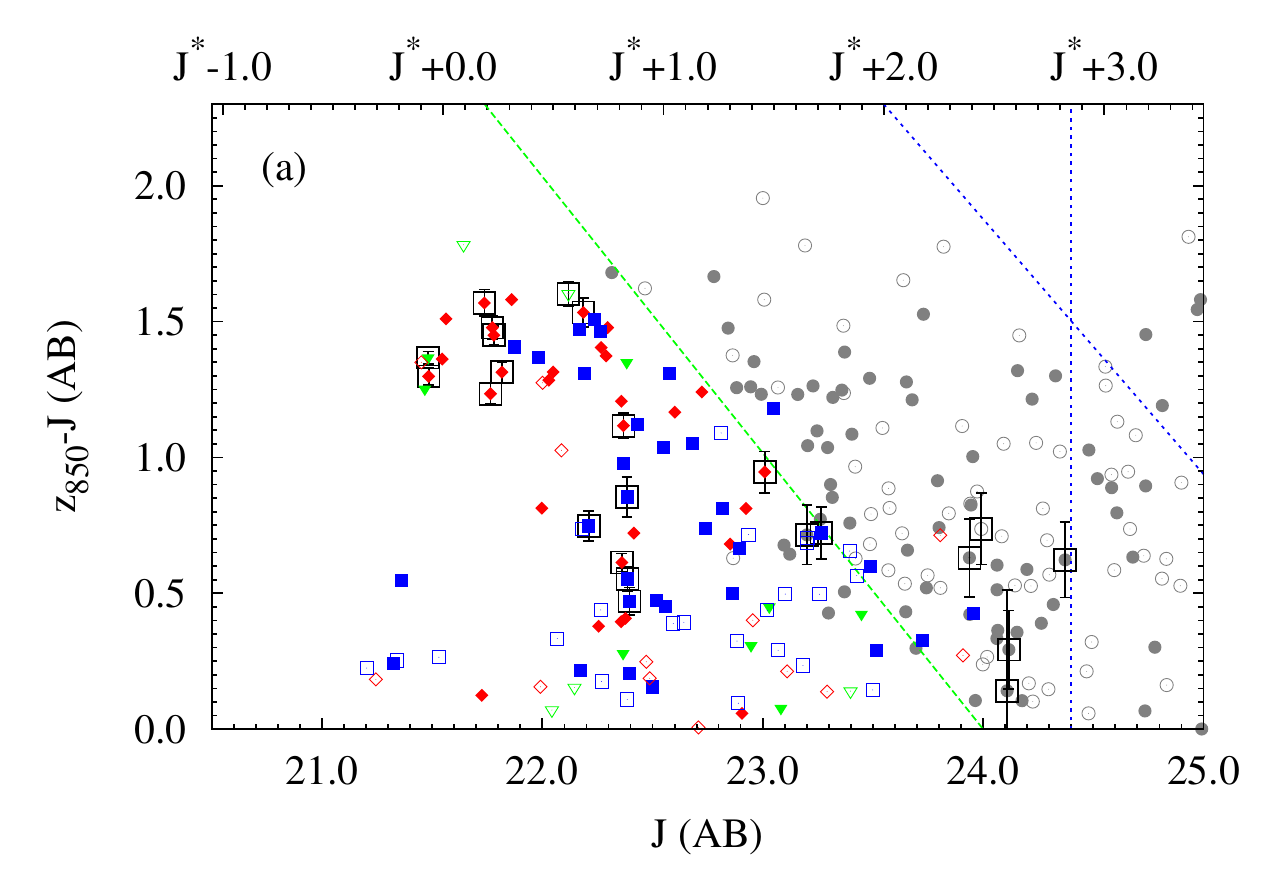}
\plotone{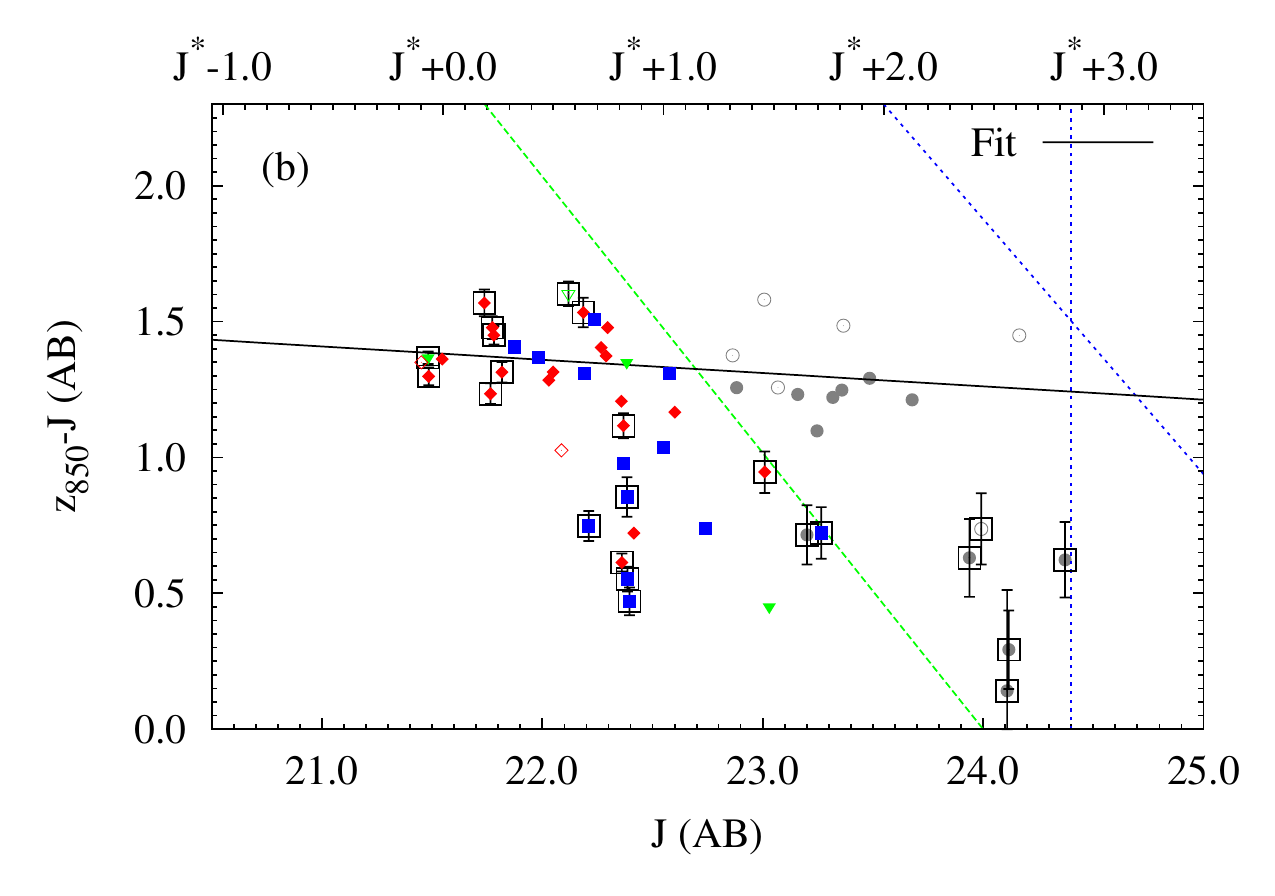}
\caption{$z_{850}-J$ versus $J$ color--magnitude diagram for J$2215.9-1738$ for (a) all galaxies; and (b) photometrically selected members only (see \textsection~\ref{s_photoZs}). Elliptical galaxies are marked as red circles; S0s with green triangles; late-type galaxies with blue squares; and galaxies for which morphologies were not determined with gray circles. Filled symbols are located within radius $r < 0.5$ Mpc of the cluster X-ray position; open symbols are located at $r > 0.5$ Mpc. Symbols enclosed within a black box are cluster members with secure spectroscopic redshifts; these points are marked with error bars, indicating the typical size of the color error in a particular region of the color--magnitude diagram. The approximate $5\sigma$ limiting magnitude for galaxies is shown by the vertical dashed blue line; the corresponding $5\sigma$ limit in $z_{850}-J$ color is shown by the diagonal dashed blue line; the diagonal dashed green line marks the approximate limit of the morphological selection. The solid black line shows the fit to the CMR for morphologically selected E+S0 galaxies (see Table~\ref{t_fitResults}).}
\label{f_zJCMD}
\end{figure}

\begin{figure}
\epsscale{0.61}
\plotone{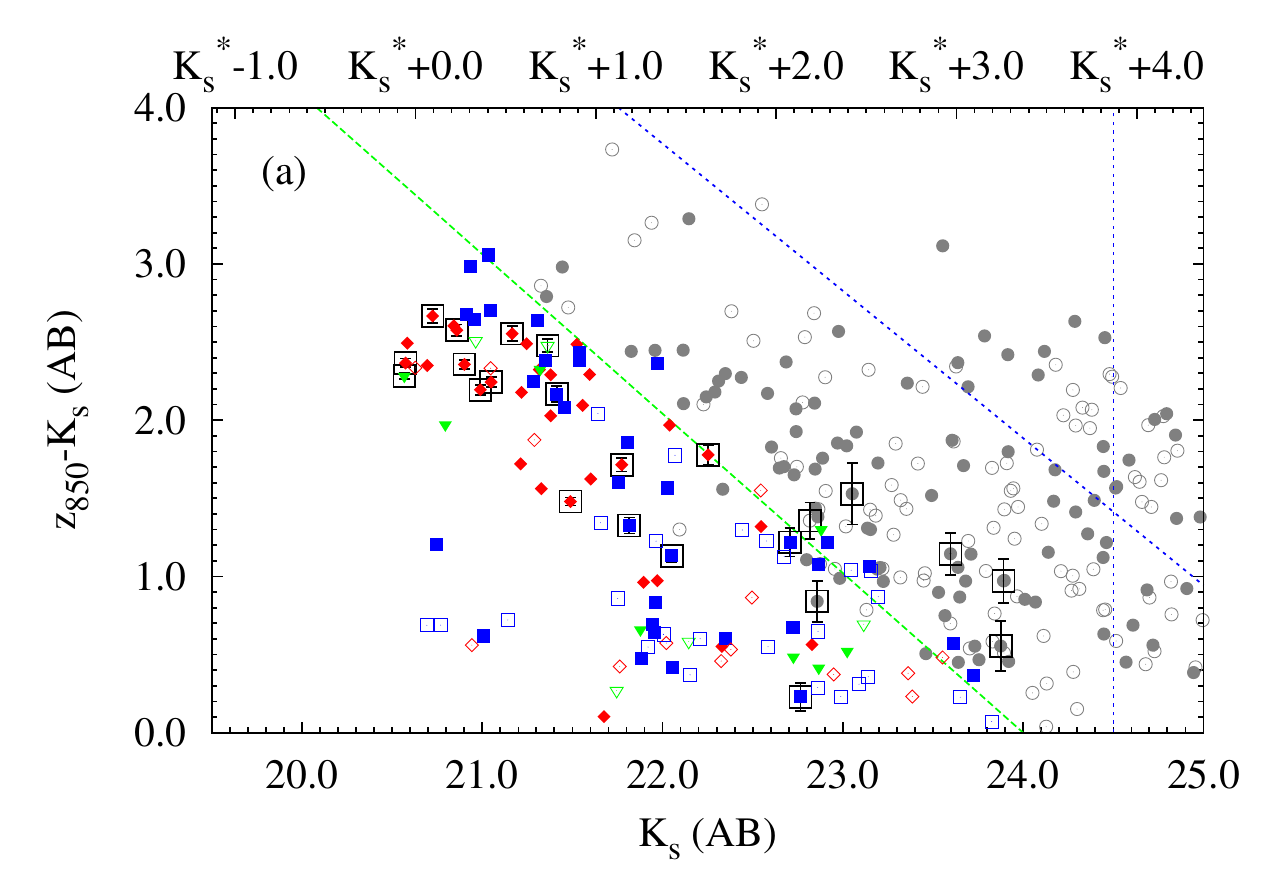}
\plotone{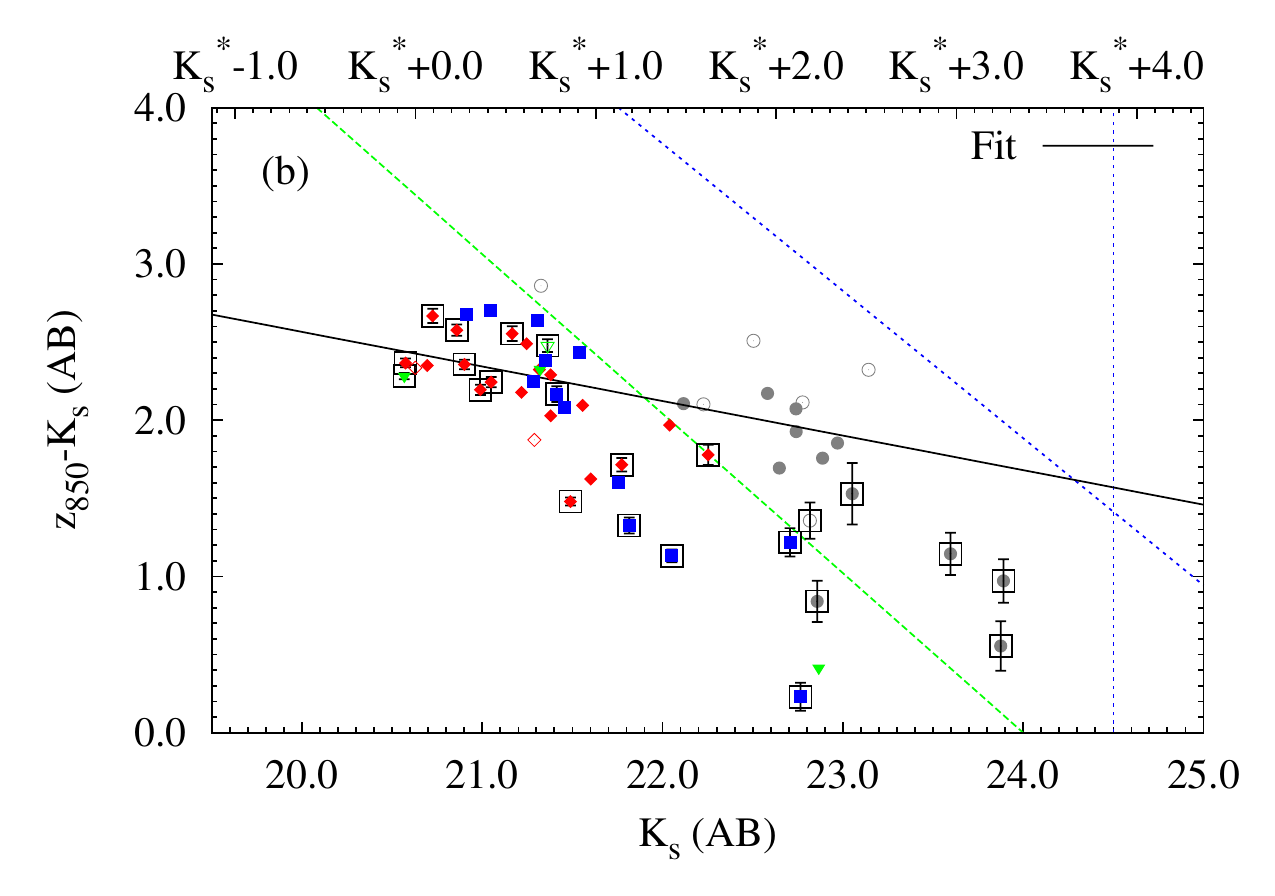}
\caption{$z_{850}-K_s$ versus $K_s$ color--magnitude diagram for J$2215.9-1738$ for (a) all galaxies; and (b) photometrically selected members only (see \textsection~\ref{s_photoZs}). The symbols and lines have the same meaning as in Fig.~\ref{f_zKCMD}. The solid black line shows the fit to the CMR for morphologically selected E+S0 galaxies (see Table~\ref{t_fitResults}).}
\label{f_zKCMD}
\end{figure}

We present the $z_{850}-J$ and $z_{850}-K_s$ color--magnitude diagrams (CMDs) with fitted color--magnitude relations (\textsection~\ref{s_resultsCMRs}, below) in Fig.~\ref{f_zJCMD} and Fig.~\ref{f_zKCMD} respectively. Panel (a) of each figure shows all galaxies in the combined ACS/MOIRCS imaging area; panel (b) shows only members selected according to the photometric criteria described in \textsection~\ref{s_photoZs} above. In each of these plots we show the approximate $5\sigma$ limiting magnitude for galaxies as a blue vertical dashed line (estimated from the size of the photometric errors); the corresponding $5\sigma$-limit in color as a diagonal dashed blue line; and the approximate limit of the morphological classification caused by the $z_{850} < 24$ limit (see \textsection~\ref{s_analysisMorphologies} above) as a diagonal dashed green line.

Along the top axis of each plot, we show the magnitude relative to the expected characteristic magnitude ($M^*$; labelled as $J^*$ or $K_s^*$ appropriately for each band in the plots) of the galaxy luminosity function at $z=1.46$. The value of $M^*$ was estimated by passively evolving the value of $K^*$ found for clusters at $z=0.9$ by \citet{DePropris_1999} to $z=1.46$, transformed to the passbands used in this paper. The evolution correction was performed by adopting a \citet{BruzualCharlot_2003} single burst of star formation model, beginning at redshift $z_f=3$, decaying exponentially with characteristic time scale $\tau = 0.1$ Gyr, with solar metallicity, and \citet{Salpeter_1955} Initial Mass Function (IMF). This model acceptably reproduces the observed evolution of $K^*$ over the redshift range $0.1 < z < 0.9$ for the data in \citet{DePropris_1999}.

It is immediately apparent from Figs.~\ref{f_zJCMD} and \ref{f_zKCMD} that the bright end of the color--magnitude relation in J$2215.9-1738$ is unpopulated: the brightest galaxies have $K_s \approx 20.6$, approximately as bright as the expected value of $M^*$. In contrast, the only other $z>1.3$ X-ray selected cluster currently known, XMMU J$2235.3-2557$ at $z=1.39$, has four spectroscopically confirmed members with $K_s$ magnitudes in the range $18.9 < K_s ({\rm AB}) < 20.6$, within a 0.2 Mpc radius of the cluster center \citep{Lidman_2008}. This corresponds to a difference in absolute magnitude of $\sim 1.5$ mag between the BCGs \citep[see also][]{Collins_2009}, neglecting the correction for passive evolution, as the difference in lookback time between the two clusters is only 0.2 Gyr.

Given the small number of systems observed to date at such high redshifts, it is unclear which of these objects is more typical of the general cluster population at this epoch. We note that the brightest galaxies in the compact cluster associated with the $z=1.51$ galaxy GDDS-12-5859 have similar $K$-band magnitudes to the brightest members of J$2215.9-1738$ \citep{McCarthy_2007}. However, further spectroscopic observations of this object are required to confirm its nature as a gravitationally bound system at $z=1.5$.

Inspection of Figs.~\ref{f_zJCMD} and \ref{f_zKCMD} shows that limiting the CMR fitting to a purely morphologically selected sample of E+S0 galaxies would restrict the total magnitude range in comparison to the magnitude limit of the complete sample. We therefore chose to perform the analysis on two samples - a morphologically selected sample of E+S0 galaxies (hereafter referred to as the `morphologically selected sample'), and a sample defined without this restriction, spanning a larger magnitude range (hereafter referred to as the `photometrically selected sample'). In both cases, we only include galaxies that pass the photometric membership selection criteria described in \textsection~\ref{s_photoZs} above, to reduce the contamination from non-cluster member galaxies significantly. For the photometrically selected sample, we applied an additional color cut of $0.8 < z_{850}-J < 1.6$, $1.5 < z_{850}-K_s < 3.0$ as appropriate for the CMR being fitted, in order to avoid biasing the fits by the inclusion of a handful of faint galaxies much bluer than the red sequence. 

An additional benefit of performing the analysis on a sample not selected on the basis of morphology is to examine the sensitivity of the results to the morphological classification. Systematic misclassification of some galaxies, for example face-on S0s as ellipticals, or flattened late type galaxies as S0s, would increase the apparent scatter about the CMR and result in the inference of younger galaxy ages \citep[see, e.g.,][]{Mei_2009}.

\subsection{Color-magnitude relations}
\label{s_resultsCMRs}
The color--magnitude relations were fitted to the color--magnitude data for each sample, defined in \textsection~\ref{s_resultsCMDs} above, using a robust biweight linear least squares method. We estimated the errors on the fits in a Monte-Carlo fashion: we generated 1,000 realizations of the data, replacing the color of each galaxy with a random variate, assuming that the color errors have a Gaussian distribution, and fitted the CMR for each of these simulated datasets. The 1$\sigma$ errors in the slope and intercept of the measured CMR were taken to be the standard deviations of the slopes and intercepts measured for the 1,000 simulated datasets.

For the morphologically selected sample of early type (E+S0) galaxies, we obtained the following relations:
\begin{align}
\label{e_morphCMRz-J}
z_{850}-J = (-0.049 \pm 0.062)&(J-22.5) \nonumber \\
&+(1.335 \pm 0.046),
\end{align}
\begin{align}
\label{e_morphCMRz-K}
z_{850}-K_s = (-0.221 \pm 0.057)&(K_s-22.5) \nonumber  \\
&+(2.012 \pm 0.091).
\end{align}

For the photometrically selected sample, we obtained similar results:
\begin{align}
\label{e_photoCMRz-J}
z_{850}-J = (-0.112 \pm 0.026)&(J-22.5) \nonumber \\
&+(1.282 \pm 0.016),
\end{align}
\begin{align}
\label{e_photoCMRz-K}
z_{850}-K_s = (-0.299 \pm 0.021)&(K_s-22.5) \nonumber  \\
&+(1.914 \pm 0.028).
\end{align}

Both the slopes and intercepts measured for the morphologically and photometrically selected samples agree at better than $<2 \sigma$ for both the fits to the $z_{850}-J$ and $z_{850}-K_s$ CMRs, with the agreement being slightly better between the two samples in $z_{850}-J$. The uncertainties on the CMR slopes and intercepts are relatively large, due to the large scatter about the CMRs that can be seen in Figs.~\ref{f_zJCMD} and \ref{f_zKCMD}.

As Figs.~\ref{f_zJCMD} and \ref{f_zKCMD} show, there are only ten spectroscopically confirmed members with early type morphologies located within 0.5 Mpc of the cluster X-ray position with $J < 22.5$. We found that fitting the CMR for this purely spectroscopically selected sample gave significantly worse constraints on the CMR than the samples supplemented with photometric redshifts; for example, the slope of the $z_{850}-J$ CMR is found to be $-0.262 \pm 0.127$. This is driven by one object (ID 526), which has a clear elliptical morphology (see Fig.~\ref{f_morphologies}), but is significantly bluer than the other spectroscopically selected early type galaxies. For this reason, we do not consider the purely spectroscopically selected galaxy sample further in this paper.

The observed scatter about the CMR was measured using a biweight scale estimate of the color residuals, with the error being estimated from 1,000 bootstrap samples. The intrinsic scatter $\sigma_{\rm int}$ in the CMR was estimated by subtracting in quadrature the RMS of the color errors of the galaxy sample to which the CMR was fitted. For the morphologically selected sample of early type (E+S0) galaxies, we measured $\sigma_{\rm int} (z_{850}-J) = 0.123 \pm 0.049$, and $\sigma_{\rm int} (z_{850}-K_s) = 0.173 \pm 0.052$. For the photometrically selected sample, we found $\sigma_{\rm int} (z_{850}-J) = 0.118 \pm 0.034$, and $\sigma_{\rm int} (z_{850}-K_s) = 0.237 \pm 0.037$. The measurements of intrinsic scatter are therefore in excellent agreement, at better than $< 1 \sigma$, between the morphologically and photometrically selected samples. This suggests that systematic morphological classification errors of the type described in \textsection~\ref{s_resultsCMDs} are not a problem at the level of precision achieveable with our galaxy sample.

We examined the dependence of our results on the photometric redshift selection in a number of ways. Adopting a conservative cut of $p_{\Delta_z} > 0.95$, we found no significant differences with the results reported above: the differences between the CMR slope and intercept values in each band and for each sample are significant at the $< 2 \sigma$ level. Restricting the photometric redshift selection to only galaxies with $z_p=1.46 \pm 0.1$, with $p_{\Delta_z} > 0.9$, we find likewise. In all cases the measurements of intrinsic scatter are robust to the changes in the CMR fit parameters; we find that the differences across all samples and in all bands are significant at the $<1 \sigma$ level.

As a final check on the robustness of the results to the photometric redshift selection, we performed 500 Monte-Carlo simulations in which, in addition to the photometry of each object in the catalog being randomized appropriately according to the size of the photometric errors, the photometric redshifts were also re-estimated using \texttt{EAZY}, and the algorithm used to select cluster members described in \textsection~\ref{s_photoZs} was applied. For each of these simulations, the CMR fit parameters and scatter measurements were performed in the manner described above. Again, we found no significant differences from the above results, with the mean slope and intercept of the Monte-Carlo simulations being within $<2 \sigma$ of the results for both the photometrically and morphologically selected samples in both $z_{850}-J$ and $z_{850}-K_s$, and the internal scatter measurements being in agreement across all samples at better than the $<1 \sigma$ level.

Table~\ref{t_fitResults} presents a summary of the measured fit parameters for the $z_{850}-J$ and $z_{850}-K_s$ CMRs, for both the morphologically and photometrically selected samples.

\subsection{Inferred ages}
\label{s_discussionScatter}
The intrinsic scatter about the CMR can be used to estimate the major epoch of star formation for early type galaxies, given the assumption of a particular stellar population model and that the CMR is predominantly a sequence in metallicity and not age \citep[e.g.][]{Bower_1992}. This latter assumption was proven to be sound by observations of the CMR in distant clusters \citep[e.g.][]{KodamaArimoto_1997}. In actual fact, recent observations have shown that the CMR is primarily a sequence in stellar mass, in that the brightest galaxies tend to be more massive, metal rich and older than the galaxies at the faint end of the CMR, with the scatter in the CMR at the faint end being primarily due to differences in age \citep[e.g.][]{Gallazzi_2006}.

We modelled the CMR scatter in the manner described by \citet[][see also \citealt{Blakeslee_2003, MeiLynx_2006}]{Bower_1992}. We assumed as our baseline galaxy model a composite stellar population (CSP) of solar metallicity, with \citet{Salpeter_1955} Initial Mass Function (IMF), formed in a single burst of star formation, decaying exponentially with characteristic time scale $\tau = 0.1$ Gyr. To check the dependence of our results on the stellar models, we used CSPs derived from \citet[][hereafter BC03, using the Padova 1994 stellar tracks]{BruzualCharlot_2003}, and \citet[][hereafter M05]{Maraston_2005}. One major difference between these two families of models is that the treatment of thermally pulsing asymptotic giant branch (TP-AGB) stars is different, with the M05 models being shown to infer younger ages for galaxies in the Hubble Ultra Deep Field (UDF) in comparison to the BC03 models \citep{Maraston_2006}.

We estimate the expected CMR scatter in the following way for each combination of CSP model and photometric bands. We assume that galaxies are born in a single burst of star formation at a time $t_{\rm birth}$ between the epoch of recombination and time $t_{\rm end}$, where the value of $t_{\rm end}$ is varied in steps between zero and the epoch at which the cluster is observed (i.e., up to a maximum of $t_z \approx 4.3$ Gyr in our adopted cosmology). For each value of $t_{\rm end}$, we construct a sample of 10,000 simulated galaxies with ages assigned randomly from a uniform distribution with range $0 < t_{\rm birth} < t_{\rm end}$, and estimate the scatter in the color distributions using a biweight scale estimate. The value of $t_{\rm end}$ corresponding to the observed intrinsic CMR scatter is then used to infer estimates of the minimum and mean luminosity weighted galaxy age assuming a particular CSP.

\begin{figure}
\epsscale{0.53}
\plotone{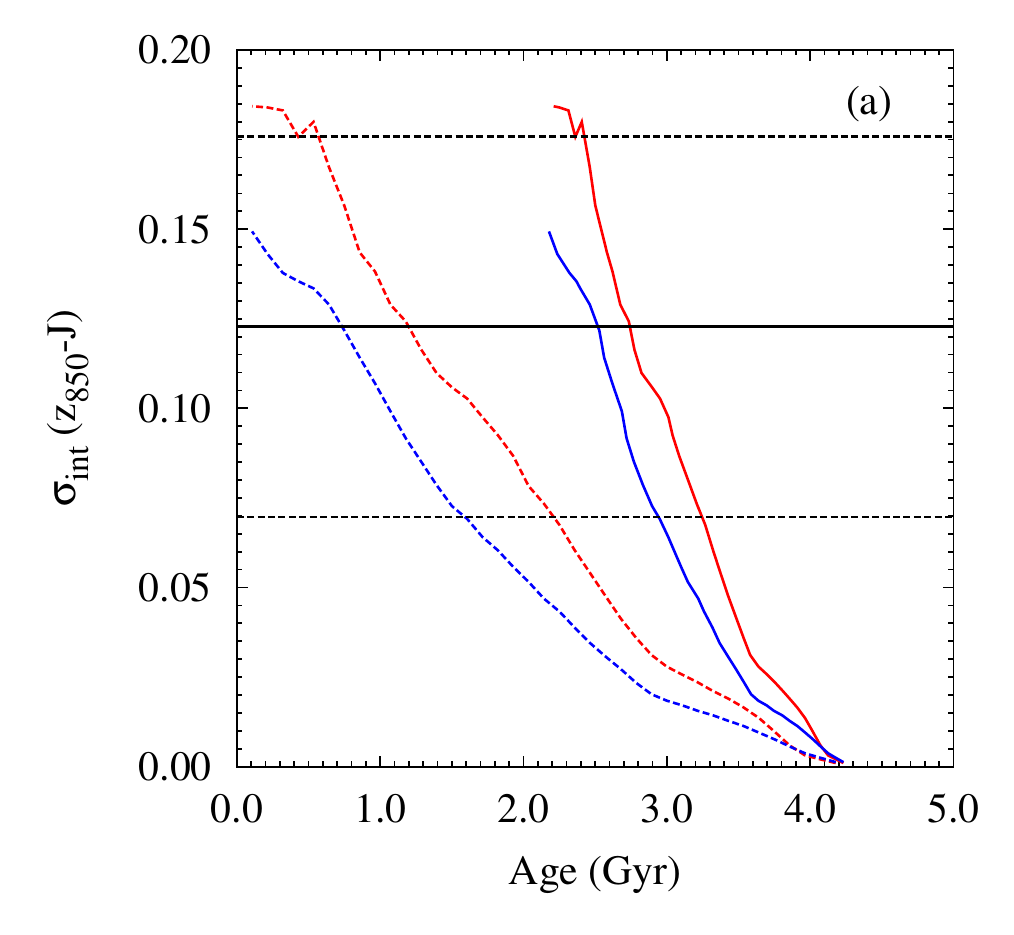}
\plotone{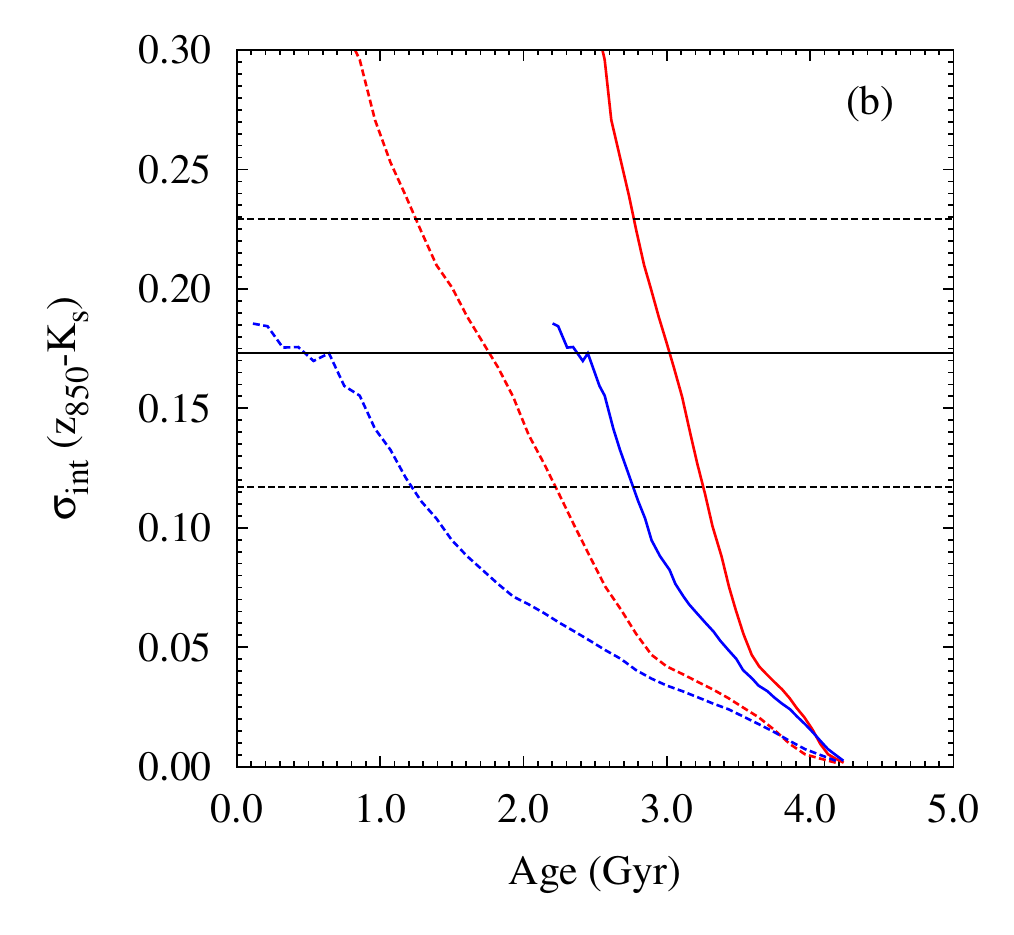}
\caption{Intrinsic scatter $\sigma_{\rm int}$ about the color--magnitude relation versus age for (a) $z_{850}-J$ and (b) $z_{850}-K_s$. The horizontal solid black line indicates the measured value of $\sigma_{\rm int}$ for the morphologically selected sample of E+S0 galaxies; the horizontal dotted lines indicate the $\pm 1\sigma$ measurement errors in $\sigma_{\rm int}$. The solid curves show the mean luminosity weighted age in the stellar population models used; the dashed curves indicate the corresponding minimum age (red: $\tau = 0.1$ Gyr burst \citet{BruzualCharlot_2003} model with solar metallicity; blue: equivalent \citet{Maraston_2005} model; see \textsection~\ref{s_discussionScatter} for details).
}
\label{f_scatterFits}
\end{figure}

Fig.~\ref{f_scatterFits} shows the expected intrinsic scatter $\sigma_{\rm int}$ around the CMR for the BC03 (red) and M05 (blue) models as a function of minimum age (dashed lines) or mean luminosity weighted age (solid lines) for the morphologically selected sample. Adopting the BC03 model, the measured intrinsic scatter around the $z_{850}-J$ CMR indicates that the minimum age of the stellar populations of the red sequence galaxies in J$2215.9-1738$ is $> 1.3 \pm 0.8$ Gyr (corresponding to a minimum redshift of formation of $z_f > 2.1^{+0.7}_{-0.5}$). The corresponding mean luminosity weighted age is $\overline{t_L} = 2.8 \pm 0.4$ Gyr (i.e., $\overline{z_f} = 4.0^{+1.1}_{-0.7}$). The ages inferred from the M05 model are younger; the minimum age being $> 0.7 \pm 1.6$ Gyr ($z_f > 1.8^{+1.4}_{-0.6}$), and the mean luminosity weighted age being $\overline{t_L} = 2.5 \pm 0.8$ Gyr ($\overline{z_f} = 3.5^{+2.3}_{-1.0}$). However, the difference between the results for the two different models is not statistically significant.

In $z_{850}-K_s$, the minimum age is $> 1.6 \pm 0.3$ Gyr ($z_f > 2.4^{+0.3}_{-0.3}$), and the mean luminosity weighted age is $\overline{t_L} = 3.0 \pm 0.2$ Gyr ($\overline{z_f} = 4.5^{+0.5}_{-0.4}$), assuming the BC03 model. This is in excellent agreement with the age inferred from the scatter about the $z_{850}-J$ CMR, though of course the colors are not completely independent. For the M05 model, we again find lower ages: the minimum age is $> 0.5 \pm 1.2$ Gyr ($z_f > 1.7^{+0.8}_{-0.5}$), and the mean luminosity weighted age is $\overline{t_L} = 2.4 \pm 0.6$ Gyr ($\overline{z_f} = 3.3^{+1.3}_{-0.7}$). 

It is interesting to note that the measured scatter about the $z_{850}-K_s$ CMR is near the limit of the allowed scatter in the M05 model (see Fig.~\ref{f_scatterFits}), although the uncertainty in the measurement is large. In fact, in the case of the photometrically selected sample, the measured scatter is larger than the maximum scatter expected in the M05 model. This could be a consequence of the greater weight given to TP-AGB stars in the M05 models relative to the BC03 models, which begins to have a significant effect at the $I$-band and redder wavelengths \citep[see Fig.~18 of][]{Maraston_2005}. The observed $K_s$-band samples the rest frame $z$-band at the redshift of the cluster.

As the measurements of the intrinsic scatter in $z_{850}-J$ and $z_{850}-K_s$ are consistent between the morphologically and photometrically selected samples (see \textsection~\ref{s_resultsCMRs} above), the inferred ages derived from the photometrically selected sample are consistent with those quoted above; however, the uncertainties are typically $\approx 30-50$\% smaller due to the increased sample size. We provide a summary of these results in Table~\ref{t_ages}.

\subsection{Dwarf-giant ratio}
\label{s_resultsDGR}
Some studies of clusters at high redshift have revealed a deficit in the number of faint red sequence galaxies in comparison to observations of clusters at low redshift \citep[e.g.][]{DeLucia_2004, DeLuciaCMR_2007, Stott_2007}. This has been attributed to `downsizing', where star formation is observed to have terminated at earlier epochs in more massive galaxies \citep{Cowie_1996}. In the case of J$2215.9-1738$,  Figs.~\ref{f_zJCMD} and \ref{f_zKCMD} show an apparent paucity of galaxies on the red sequence at faint magnitudes ($J > 23.5$), which may be indicative of this phenomenon.

We quantify this deficit using the Dwarf-to-Giant Ratio (DGR). Following \citet{DeLuciaCMR_2007}, we define giants as galaxies with absolute $V$-band magnitudes brighter than $M_V ({\rm Vega}) = -20$, and dwarfs as galaxies fainter than this limit, but brighter than $M_V = ({\rm Vega}) = -18.2$. These limits correspond to values at $z=0$ after correction for the effect of passive evolution. In performing the passive evolution correction, we adopt the same \citet{BruzualCharlot_2003} simple stellar population model as used by \citet{DeLuciaCMR_2007}, with solar metallicity, \citet{Chabrier_2003} IMF, and formation redshift $z_f = 3$. The dividing magnitude between dwarfs and giants is equivalent to $J ({\rm AB}) \approx 22.7$ in the observed frame of J$2215.9-1738$ after this correction is applied, and the corresponding faint magnitude limit is $J ({\rm AB}) \approx 24.5$ (see \textsection~\ref{s_discussionU-VScatter} below for details of the adopted conversion between $V$ and $J$ magnitudes). Thus, the faint magnitude limit is $\approx 0.1$ mag. fainter than the 5$\sigma$ limiting magnitude of our observations, and so the limits that we obtain on the DGR may be slightly underestimated in comparison to \citet{DeLuciaCMR_2007}.

We estimated the DGR using the sample of photometrically selected cluster members defined in \textsection~\ref{s_photoZs} and listed in Table~\ref{t_galaxyPhotometry}. Following \citet{DeLuciaCMR_2007}, using all of the members within $r < 0.5 R_{200}$ ($R_{200} \approx 0.6$ Mpc for J$2215.9-1738$; \citealt{Hilton_2007}), we find that the ${\rm DGR}=0.50 \pm 0.21$. In estimating the DGR of galaxies on the red sequence, we only consider the $z_{850}-J$ CMR, as unlike the $z_{850}-K_s$ CMR, it is well suited to comparison with other studies at low redshift. Applying a color cut of $\pm0.3$ mag. around the fit to the $z_{850}-J$ CMR derived from the morphologically selected sample, we find that the red sequence ${\rm DGR} = 0.32\pm 0.18$. The uncertainties are estimated assuming Poissonian errors, added in quadrature to the uncertainty due to the photometric redshift selection, calculated using the full \texttt{EAZY} photometric redshift probability distribution for each galaxy by the same technique described in \textsection~\ref{s_resultsMorphologies} above. The uncertainty due to the photometric redshift selection typically increases the size of the error bars on the DGR measurements by $\approx 12$\%.

We checked the sensitivity of these results to the parameters governing the photometric redshift selection. Restricting the membership selection to galaxies with $p_{\Delta_z} >0.95$ or $1.36 < z_p < 1.56$ does not affect these results significantly, the DGR measurements in each case being well within $<1 \sigma$ of the above results.

Extending the selection radius to $r < 0.75$ Mpc does not change these results significantly: we estimate that the ${\rm DGR}=0.65 \pm 0.22$ if all the photometrically selected members are included; if we include only those members within $\pm 0.3$ mag. of the red sequence, then we find that the ${\rm DGR}=0.46 \pm 0.20$.

\section{Discussion}
\label{s_discussion}

\subsection{Morphologies}
The morphological fractions we find for J$2215.9-1738$ are in good agreement with studies of clusters at $z \sim 1$, some $\sim 1.5$ Gyr later than the epoch at which our target cluster is observed.

\citet{Smith_2005} studied the morphology--density relation in a sample of six clusters at $z \sim 1$. Assuming that $J^*+1$ is roughly equivalent to the magnitude limit of $M_{V}^*+1$ adopted by \citet{Smith_2005}, as at the redshift of J$2215.9-1738$ the $J$-band samples the rest frame $V$-band, then the depth of our study is sufficient to make a comparison, as all the galaxies in our sample brighter than this limit have morphological classifications (see Fig.~\ref{f_zJCMD}). As \citet{Smith_2005} quote their results in bins of local galaxy surface density $\Sigma$, we need to estimate the median local galaxy surface density for our sample in order to compare results. Following \citet{Smith_2005}, we calculate $\Sigma$ for each galaxy by counting the 10 nearest neighbors with $J < J^*+1$, and dividing by the rectangular area enclosed. We find the median local galaxy surface density within a radius of 0.5 Mpc of the cluster X-ray position is $\overline{\Sigma} \approx 227$ Mpc$^{-2}$. In their second highest density bin with $\Sigma > 100$ Mpc$^{-2}$, \citet{Smith_2005} find $f_{E+S0} = 0.6 \pm 0.1$, in good agreement with the results of our study.

A study of the morphology--density relation at $z \sim 1$ was also conducted by the ACS Guaranteed Time Observations (GTO) team \citep{Postman_2005}, who quote morphological fractions within $R_{200}$ for the individual objects that make up their sample. The magnitude limit of the \citet{Postman_2005} study is $z_{850} \leq 24$, as adopted in this work. For J$2215.9-1738$, the $R_{200}$ radius is $\approx 0.6$ Mpc \citep{Hilton_2007}, i.e. approximately the same as the 0.5 Mpc radius in which we have calculated the morphological fractions. The mean morphological fractions within $R_{200}$ for the \citet{Postman_2005} sample ($\pm$ the standard error on the mean) are $f_E = 0.38 \pm 0.03$; $f_{S0} = 0.16 \pm 0.04$; $f_{Sp+Irr} = 0.48 \pm 0.07$; and $f_{E+S0} = 0.55 \pm 0.07$. The results we obtain for J$2215.9-1738$ are in agreement at the $< 1 \sigma$ level for all morphological bins. A larger cluster sample would be needed to search for possible evolutionary effects between $z \sim 1.5$ and $z \sim 1$.

\subsection{Inferred ages}
The mean redshift we find for the major epoch of star formation in the early type galaxies in J$2215.9-1738$ of $z_f \approx 3-5$ (see \textsection~\ref{s_discussionScatter}) is similar to that found for elliptical galaxies in other high redshift clusters. \citet{MeiLynx_2006} found a mean formation redshift of $z_f \approx 4$ for the elliptical galaxies within 0.5 Mpc radius of the centers of the two Lynx clusters at $z=1.26$, from modelling the CMR scatter using single burst BC03 models, in good agreement with our result. Similarly, \citet{Lidman_2008} conducted a study of the $J-K_s$ CMR in the only other $z>1.3$ X-ray selected cluster currently known, XMMU J$2235.3-2557$, finding $z_f \approx 4$ for galaxies, selected using a color cut, within 0.2 Mpc of the cluster center.

For well studied clusters at lower redshift, slightly smaller formation redshifts have been measured from the scatter about the CMR: \citet{Blakeslee_2003} estimated $z_f \approx 2.7$ from the $i_{775}-z_{850}$ E+S0 CMR of RDCS J$1252.9-2927$ at $z=1.24$, a result also found by \citet{Lidman_2004} from the $J-K_s$ CMR; \citet{MeiRDCS_2006} found $z_f \approx 3$ from the CMR of elliptical galaxies in RDCS J$0910+5422$ at $z=1.10$; and \citet{Blakeslee_2006} estimate that the elliptical galaxies in the $z=0.83$ clusters MS $1054-03$ and RX J$0152.7-1357$ completed their major epoch of star formation at $z_f > 2.2$. All of these studies used BC03 stellar population synthesis models in arriving at their results.

\subsection{Evolution of the color--magnitude relation}
\label{s_discussionU-VScatter}
To compare our measurements of color--magnitude relation parameters with other studies at lower redshift, we derived linear conversions between observed colors and magnitudes and $U-V$, $V$ (Vega) in the rest frame of the Coma cluster, using $\tau =0.1$ Gyr \citet{BruzualCharlot_2003} models of several metallicities ($0.2Z_\sun$, $0.4Z_\sun$, $Z_\sun$, $2.5Z_\sun$), with a range of ages (corresponding to $2 < z_f < 7$, as appropriate for the early type population, \textsection~\ref{s_discussionScatter} above). The method used is the same as that described in Appendix II of \citet{Mei_2009}, except that we convert magnitudes to $V$ apparent (Vega) at the distance of the Coma cluster. The following transformations were used:

\begin{align}
\label{e_zJUV}
(U-V)_{z=0.02} = (1.178 \pm 0.003)&(z_{850}-J) \nonumber \\
&-(0.658 \pm 0.005),
\end{align}

\begin{align}
\label{e_JV}
V_{z=0.02} = J-(0.098 \pm 0.003)&(z_{850}-J) \nonumber \\
&-(9.455 \pm 0.004).
\end{align}

Note that in the following discussion, we only consider the $z_{850}-J$ CMR of J$2215.9-1738$, as the $z_{850}-K_s$ CMR (roughly equivalent to rest frame $U-z$) is a poor match to other studies in the literature. As the CMR fitting parameters for J$2215.9-1738$ derived from the morphologically and photometrically selected samples are consistent, we only quote transformed quantities corresponding to the morphologically selected sample below.

\begin{figure}
\epsscale{0.75}
\plotone{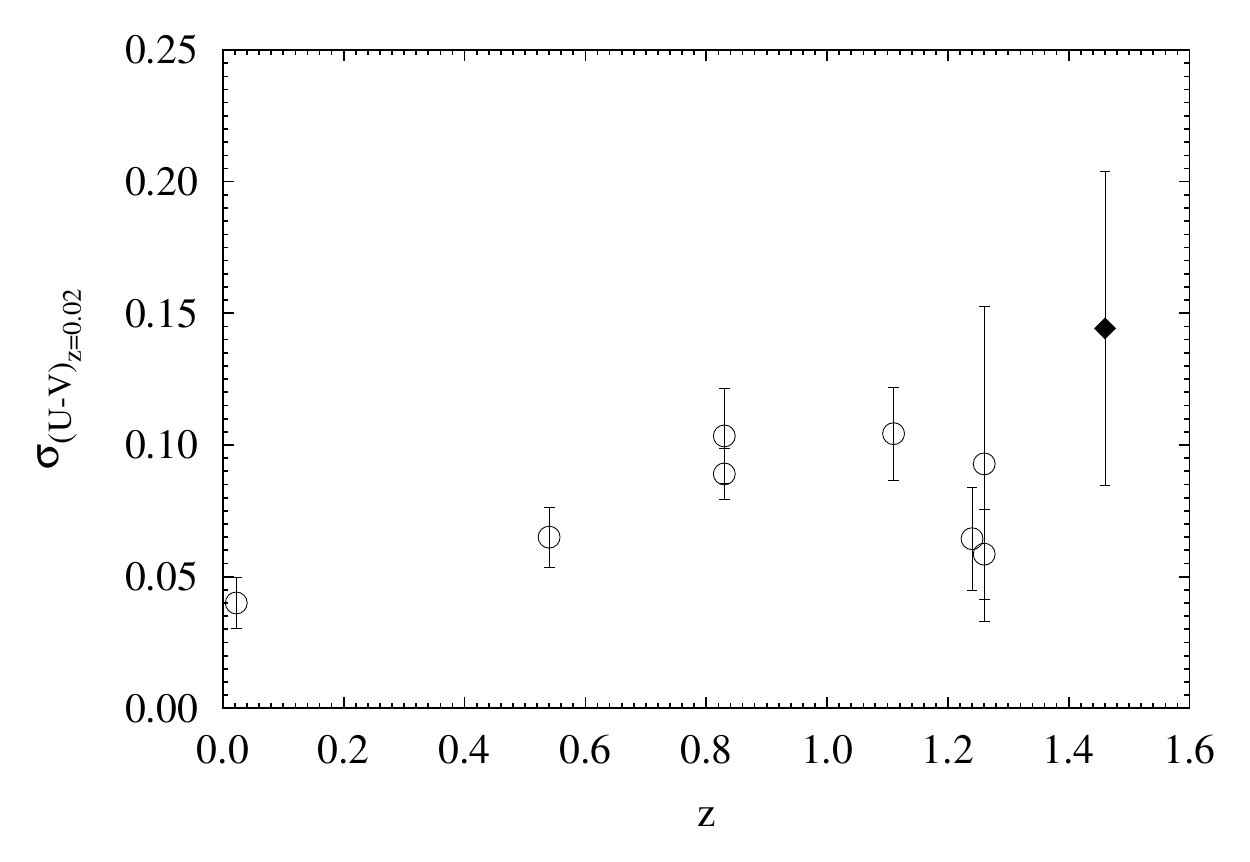}
\caption{Evolution of the intrinsic scatter around the color--magnitude relation with redshift in $U-V$ in the rest frame of the Coma cluster (converted from the observed frame using \citet{BruzualCharlot_2003} models). Open points show values for several clusters in the literature from the works of \citet{Bower_1992}, \citet{Ellis_1997}, \citet{Blakeslee_2006} (two $z=0.83$ clusters), \citet{MeiRDCS_2006}, \citet{Blakeslee_2003}, \citet{MeiLynx_2006} (two $z=1.26$ clusters), ordered by increasing redshift. The black diamond marks the value derived from the $z_{850}-J$ CMR of J$2215.9-1738$ (morphologically selected sample of E+S0 galaxies). The scatter is increasing with redshift, as expected for the passive evolution of a stellar population formed at much higher redshift.
}
\label{f_U-VScatterEvolution}
\end{figure}

Fig.~\ref{f_U-VScatterEvolution} shows the evolution of the intrinsic scatter $\sigma_{\rm int}(U-V)_{z=0.02}$ about the CMR, as traced by studies of several clusters at different redshifts. Some of the studies shown in this plot measured $\sigma_{\rm int}$ through passbands roughly equivalent to $U-V$ at their respective redshifts, and so the correction to $(U-V)_{z=0.02}$ using the models is small. These are the \citet{Bower_1992} study of the Coma cluster; the work by \citet{Ellis_1997} at $z\approx 0.54$; and the two $z=0.83$ clusters observed by \citet{Blakeslee_2006}.

At $z>1$, the ACS Guaranteed Time Observations (GTO) team measured $\sigma_{\rm int}$ in $i_{775}-z_{850}$, which is more closely matched to rest frame $U-B$, and so the transformation of these results to $(U-V)_{z=0.02}$ is larger and more uncertain. Both the \citet{Blakeslee_2003} study of RDCS J$1252.9-2927$ at $z=1.24$, and the \citet{MeiLynx_2006} study of the Lynx clusters at $z=1.26$, show similar values of $\sigma_{\rm int}$ to those measured at $z=0.54$ by \citet{Ellis_1997}, and $z=0.83$ by \citet{Blakeslee_2006}, although the uncertainties are large.

The general trend in Fig.~\ref{f_U-VScatterEvolution} is of increasing scatter about the CMR as redshift increases, as expected if the stellar populations were formed at much higher redshifts than observed and have evolved passively thereafter. The intrinsic scatter about the red sequence observed in J$2215.9-1738$ at $z=1.46$ is $(U-V)_{z=0.02} = 0.144 \pm 0.059$, more than three times as large as that observed in the Coma cluster \citep{Bower_1992, Eisenhardt_2007}, some $\sim 9$ Gyr earlier, and one and a half times that observed in MS $1054-03$ and RX J$0152.7-1357$ at $z=0.83$, $\sim 2.2$ Gyr earlier.

\begin{figure}
\epsscale{0.75}
\plotone{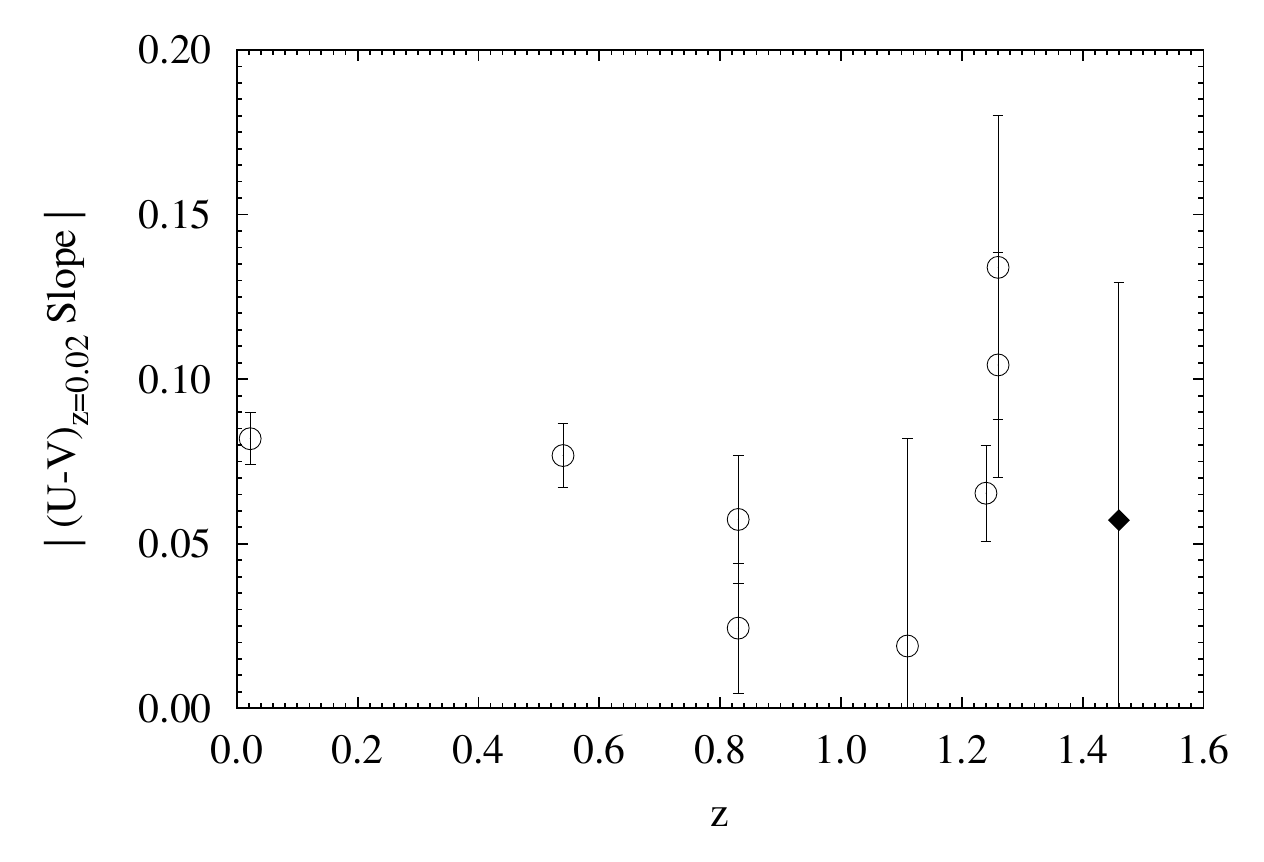}
\caption{Evolution of the absolute slope of the $U-V$ color--magnitude relation in the rest frame of the Coma cluster (converted from the observed frame using \citet{BruzualCharlot_2003} models). Open points show the values for the same comparison sample of clusters drawn from the literature as shown in Fig.~\ref{f_U-VScatterEvolution}. The black diamond marks the value derived from the $z_{850}-J$ CMR of J$2215.9-1738$ (morphologically selected sample of E+S0 galaxies). There is no evidence for evolution of the value of the slope with redshift, though the uncertainties in the slope measurements at high redshift are very large.
}
\label{f_U-VSlopeEvolution}
\end{figure}

Fig.~\ref{f_U-VSlopeEvolution} shows values of the absolute $(U-V)_{z=0.02}$ CMR slope for the same comparison sample of clusters from the literature. In contrast to the measurements of intrinsic scatter about the CMR, we see no evidence for evolution in the slope over the last $\sim 9$ Gyr. For J$2215.9-1738$, the value of the CMR slope is $-0.057 \pm 0.072$, after conversion to $(U-V)_{z=0.02}$. This agrees with the slope measured for the Coma cluster \citep{Bower_1992, Eisenhardt_2007} at better than the $<1 \sigma$ level, though the uncertainty is large. The lack of evolution of the CMR slope is consistent with the CMR remaining as primarily a sequence in metallicity up to the highest redshifts so far observed.

To compare the evolution of the CMR intercept, we used the CMR of Coma as measured by \citet{Bower_1992} to calibrate the CMR as a metallicity sequence, and transformed the CMRs of J$2215.9-1738$ and the literature comparison sample, such that the intercept was evaluated at the $V$-band magnitude corresponding to solar metallicity. The metallicity calibration of the CMR of Coma was performed by calculating $V$-band (Vega) magnitudes corresponding to the $U-V$ colors of $\tau= 0.1$ Gyr BC03 models of several metallicities, assuming the slope and intercept of the Coma CMR as measured by \citet{Bower_1992}, and formation redshift $z_f=2$ \citep{Bower_1992}. Under these assumptions, solar metallicity in the Coma CMR corresponds to $V ({\rm Vega}) = 13.27$. We calculated the corresponding $V$-band magnitudes and transformed CMR intercepts for the other clusters in the comparison sample assuming passive evolution of this model. Errors in the transformed values were estimated using a Monte-Carlo method, taking into account the uncertainties in the measured CMRs, and the color and magnitude transformations to $U-V_{z=0.02}$, $V$ (Vega).

\begin{figure}
\epsscale{0.75}
\plotone{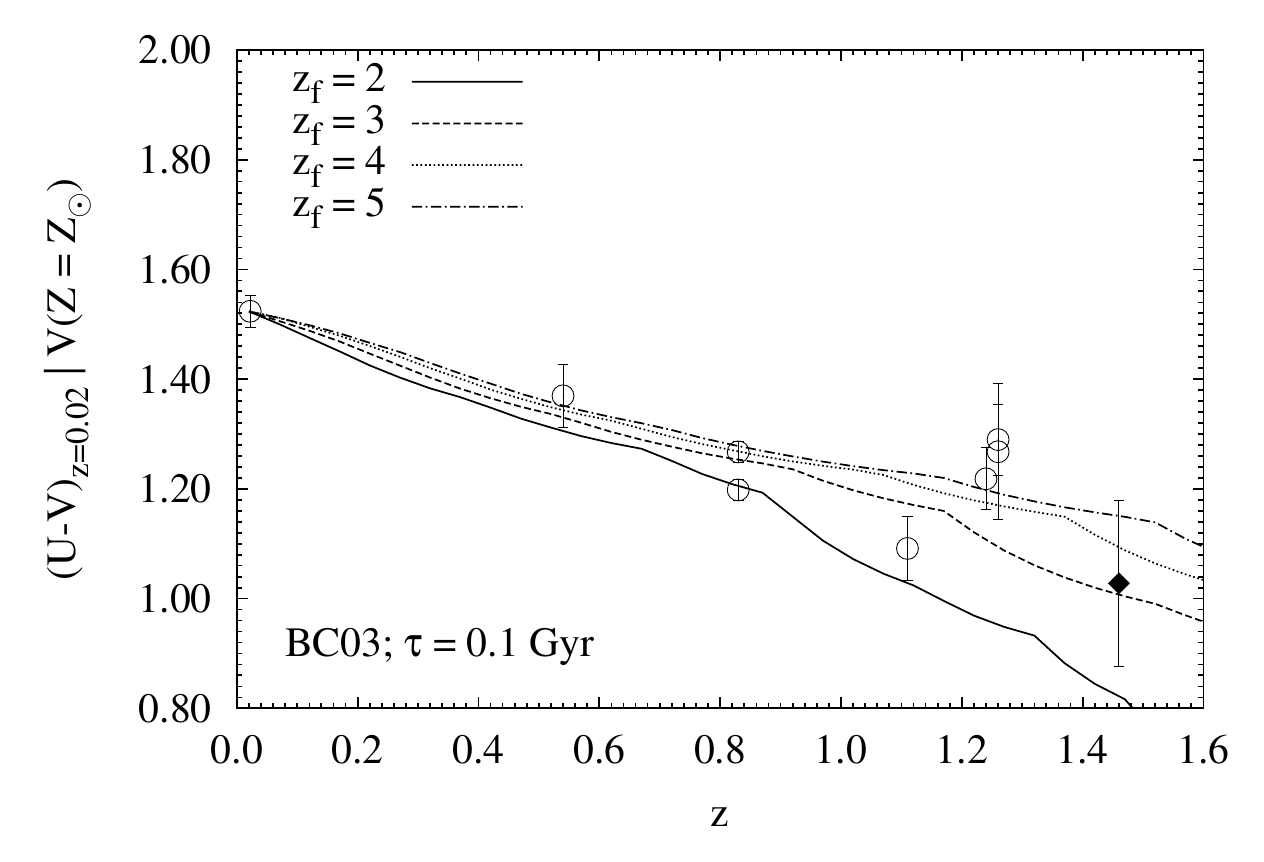}
\caption{Evolution of the expected $U-V$ color of a galaxy with solar metallicity, in the rest frame of the Coma cluster (converted from the observed frame using \citet{BruzualCharlot_2003} models), where the Coma CMR has been used to calibrate the CMR as a magnitude-metallicity relation, under the assumption that $z_f=2$ for Coma. Passive evolution has been taken into account in transforming the CMRs from the literature assuming the same $z_f=2$ model. The expected passive evolution tracks for several other formation redshifts are also plotted. Open points show the values for the same comparison sample of clusters drawn from the literature as shown in Fig.~\ref{f_U-VScatterEvolution}. The black diamond marks the value derived from the $z_{850}-J$ CMR of J$2215.9-1738$ (morphologically selected sample of E+S0 galaxies). The scatter of the data points about the model evolution tracks may indictate variation in the star formation histories of the galaxy populations between clusters.
}
\label{f_U-VSolarMetallicityEvolution}
\end{figure}

Fig.~\ref{f_U-VSolarMetallicityEvolution} shows the results of this exercise, with the expected passive evolution track of the baseline stellar population model overplotted, for several different formation redshifts. The CMR intercept evaluated at solar metallicity for J$2215.9-1738$ is consistent with $z_f =3$, which is slightly lower than the formation redshift inferred from the intrinsic scatter (\textsection~\ref{s_discussionScatter}, above), but in good agreement given the large uncertainties. This is consistent with the literature comparison sample: there is some scatter, which could indicate a range of star formation histories for the galaxy population in different clusters, but the uncertainties are large. Photometric calibration uncertainties may also contribute to the scatter; see, for example, the discussion in the \citet{MeiLynx_2006} study of the Lynx clusters at $z=1.26$.

\begin{figure}
\epsscale{0.75}
\plotone{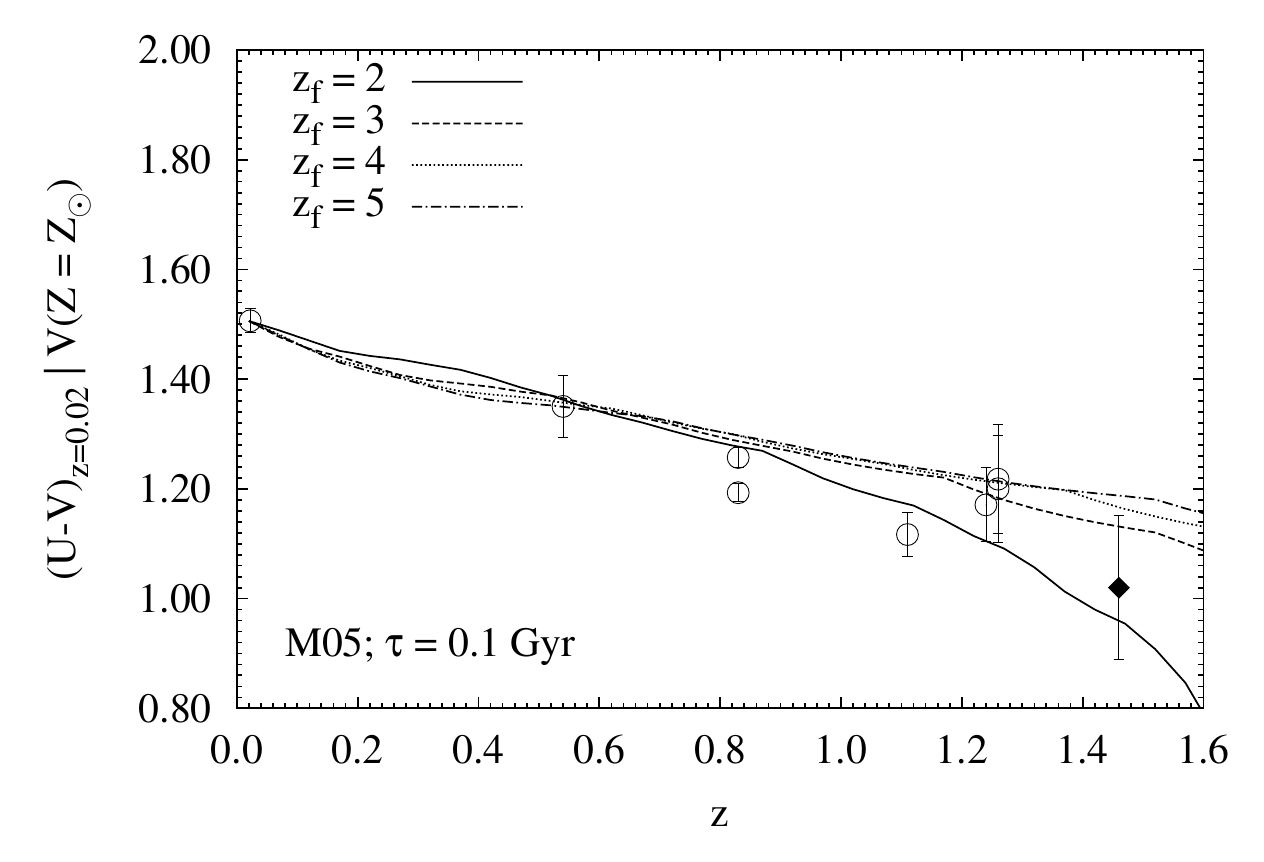}
\caption{As Fig.~\ref{f_U-VSolarMetallicityEvolution}, except the expected evolution of the $U-V$ color at solar metallicity is calculated from the passive evolution of \citet{Maraston_2005} $\tau=0.1$ Gyr burst models.
}
\label{f_U-VSolarMetallicityEvolutionMaraston}
\end{figure}

We show the effect of changing the stellar population model from BC03 to M05 on the expected passive evolution of the colors of solar metallicity galaxies in Fig.~\ref{f_U-VSolarMetallicityEvolutionMaraston}. Lower formation redshifts are expected if this family of models is assumed, with some of the literature comparison sample being expected to have formed most of their stars at $z < 2$. In the case of J$2215.9-1738$, Fig.~\ref{f_U-VSolarMetallicityEvolutionMaraston} implies $z_f \approx 2$. Again, this is in good agreement with the age derived from the intrinsic scatter about the CMR for the M05 models.

\subsection{Dwarf-giant ratio}
\label{s_discussDGR}
The Dwarf-to-Giant Ratio measured for J$2215.9-1738$ appears to be significantly lower than has been measured in clusters at lower redshifts. \citet{DeLuciaCMR_2007} quote a red sequence luminous-to-faint ratio within $r < 0.5R_{200}$ for the Coma cluster equivalent to a DGR of $3.13 \pm 0.59$, nearly ten times the value of the red sequence DGR in J$2215.9-1738$ ($0.32\pm 0.18$). Although the DGR we measure is likely to be slightly underestimated, given that our magnitude limit is $\approx 0.1$ mag brighter than the faint magnitude limit adopted by \citet{DeLuciaCMR_2007}, it seems very unlikely that this would be enough to account for the deficit of faint red sequence galaxies relative to Coma.

Several studies at lower redshift have parametrized the evolution of the red sequence DGR with redshift. Recently, \citet{GilbankBalogh_2008} compiled several measurements of the DGR in clusters at $z<1$, including studies by \citet*{Barkhouse_2007}; \citet{Gilbank_2008, Hansen_2007, DeLuciaCMR_2007}; and \citet{Stott_2007}, transformed on to a common system defined by \citet{DeLuciaCMR_2007}, as we have adopted in performing our estimate of the DGR (described in \textsection~\ref{s_resultsDGR} above). \citet{GilbankBalogh_2008} found that the evolution of the DGR with redshift of their composite sample is $\propto (1+z)^{-1.8\pm0.5}$. At the redshift of J$2215.9-1738$, the expected ${\rm DGR} = 0.68 \pm 0.36$ from this relation, in good agreement with our estimate of the DGR.

We conclude that J$2215.9-1738$ appears to show a significant deficit of faint red sequence galaxies compared to clusters in the local universe, and that the measured DGR is of similar magnitude to that expected from extrapolation of the DGR-redshift relation measured at $z < 1$. However, it is known that the cluster-to-cluster scatter of the DGR is considerable \citep{DeLuciaCMR_2007}, and study of a larger sample of clusters, ideally with large numbers of spectroscopically confirmed members, at similar redshift to J$2215.9-1738$ is required in order to confirm whether or not this holds for the general population at this epoch.

\section{Conclusions}
\label{s_conclusions}
We have conducted a study of the morphological fractions and color--magnitude relation in the galaxy cluster XMMXCS J$2215.9-1738$ at $z=1.46$. This is the first such study of an X-ray selected cluster at $z\sim1.5$. We found:

\begin{enumerate}
\item{The brightest members of J$2215.9-1738$ have $K_s$ magnitudes corresponding to the expected value of $\sim M^*$ in the galaxy luminosity function. This is significantly fainter than the brightest galaxies in the only other $z>1.3$ X-ray selected cluster studied to date, XMMU J$2235.3-2557$ at $z=1.39$ \citep{Lidman_2008}.}

\item{The morphological fractions are $f_{E} = 0.54 \pm 0.17$, $f_{S0} = 0.08 \pm 0.07$, $f_{Sp+Irr} = 0.38 \pm 0.15$, similar to other clusters at $z \sim 1$. Thus, the dominant component of the galaxy population observed in clusters at low redshift was already in place $\sim 4.5$ Gyr after the Big Bang.}

\item{After transformation from $z_{850}-J$ to $(U-V)_{z=0.02}$, the slope of the color--magnitude relation is consistent with that of the Coma cluster, implying little evolution over the last $\sim 9$ Gyr, though the measurement uncertainty is large. In contrast, the intrinsic scatter about the color--magnitude relation is more than three times the value measured in the Coma cluster, after conversion from $z_{850}-J$ to $(U-V)_{z=0.02}$.}

\item{From comparison with stellar population models, the intrinsic scatter about the color--magnitude relation implies mean luminosity weighted ages for the stellar populations of the early type galaxies in J$2215.9-1738$ of $\approx 3$ Gyr, corresponding to the main epoch of star formation in these galaxies coming to an end at $z_f \approx 3-5$. Comparison of the intercept of the color--magnitude relation with passive evolution of the same stellar population models, calibrated relative to the Coma cluster, yields consistent results.}

\item{J$2215.9-1738$ shows evidence of the `downsizing' phenomenon: the red sequence Dwarf-to-Giant ratio for the cluster is $0.32\pm 0.18$ within a radius of $0.5R_{200}$, although this is likely to be underestimated slightly in comparison to other studies due to the depth of our photometry. This is consistent with extrapolation of the redshift evolution of the DGR measured from cluster samples at $z < 1$ within the large uncertainties.}

\end{enumerate}

\acknowledgments

We thank the referee for a number of suggestions that improved the clarity of this paper. This work is based in part on data collected at the Subaru Telescope, which is operated by the National Observatory of Japan, and \textit{XMM}-Newton, an ESA science mission funded by contributions from ESA member states and from NASA. We thank Ichi Tanaka for the development of the \texttt{MCSRED} package used to reduce the MOIRCS data. We acknowledge financial support from the South African National Research Foundation, the UK Science and Technology Facilities Council, and the University of Sussex Physics \& Astronomy Department. Financial support for this work was also provided by NASA through program GO-10496 from the Space Telescope Science Institute, which is operated by AURA, Inc., under NASA contract NAS 5-26555. This work was also supported in part by the Director, Office of Science, Office of High Energy and Nuclear Physics, of the U.S. Department of Energy under Contract No. AC02-05CH11231, as well as a JSPS core-to-core program ``International Research Network for Dark Energy'' and by JSPS research grant 20040003. This work was performed under the auspices of the U.S. Department of Energy by Lawrence Livermore National Laboratory in part under Contract W-7405-Eng-48 and in part under Contract DE-AC52-07NA27344. The authors wish to recognize and acknowledge the very significant cultural role and reverence that the summit of Mauna Kea has always had within the indigenous Hawaiian community; we are fortunate to have the opportunity to conduct observations from this mountain.

{\it Facilities:} \facility{HST}, \facility{Keck:II}, \facility{Subaru}, \facility{VLT:Antu}, \facility{XMM}


\clearpage

\clearpage

\begin{deluxetable}{c}
\includegraphics[height=230mm]{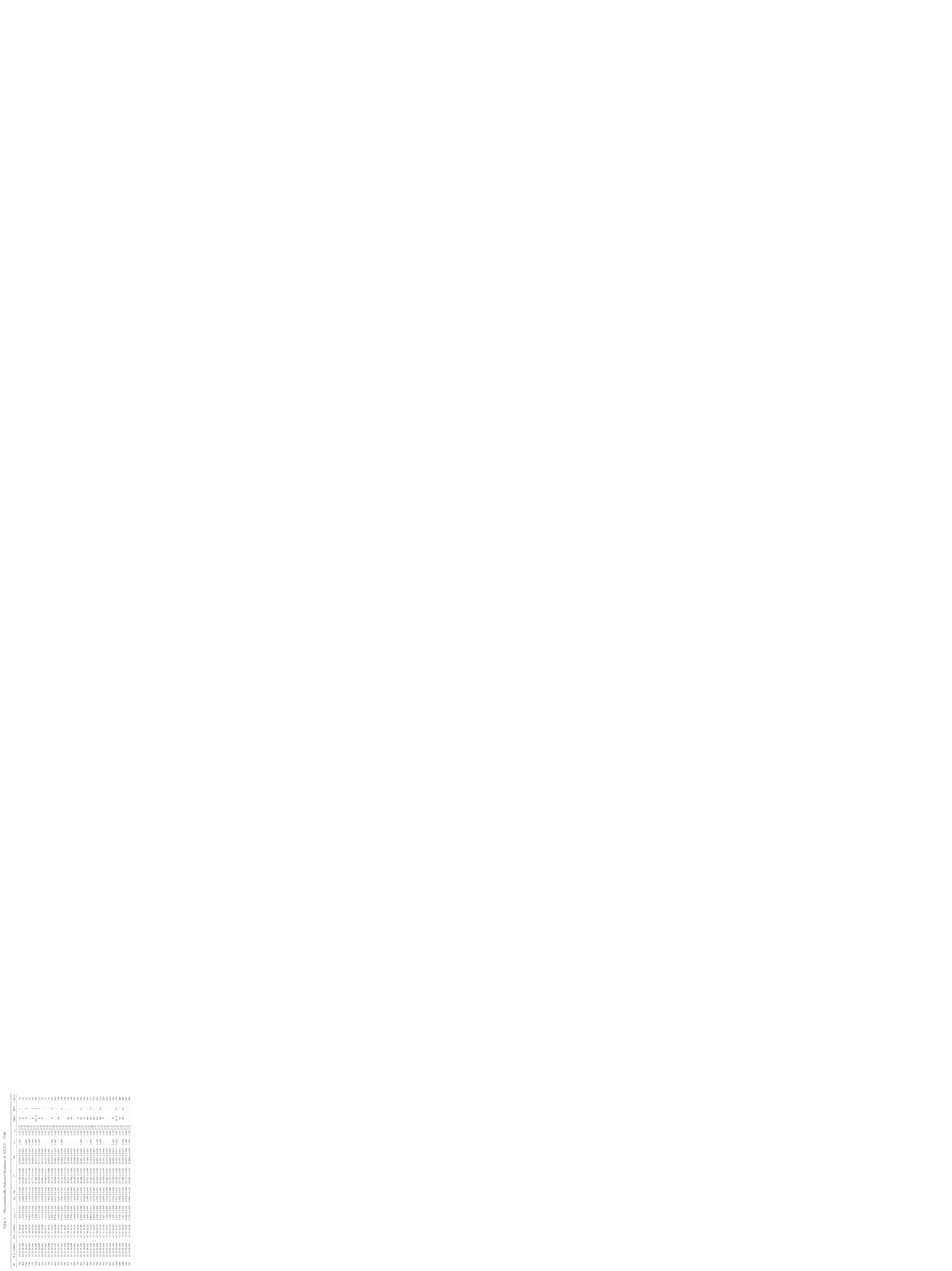}
\label{t_galaxyPhotometry}
\end{deluxetable}

\clearpage

\begin{figure}
\includegraphics[height=230mm]{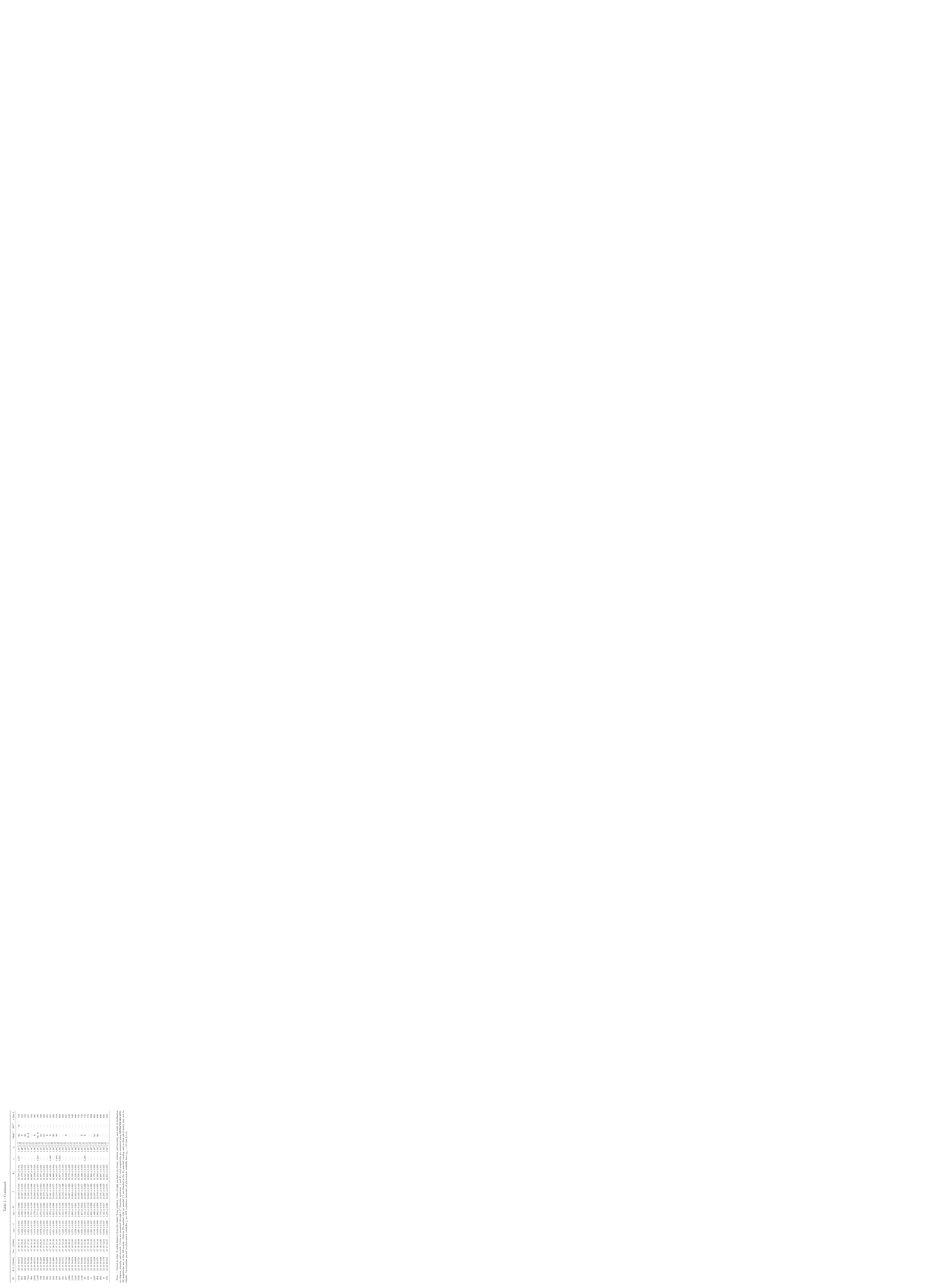}
\end{figure}

\clearpage

\begin{figure}
\includegraphics[height=230mm]{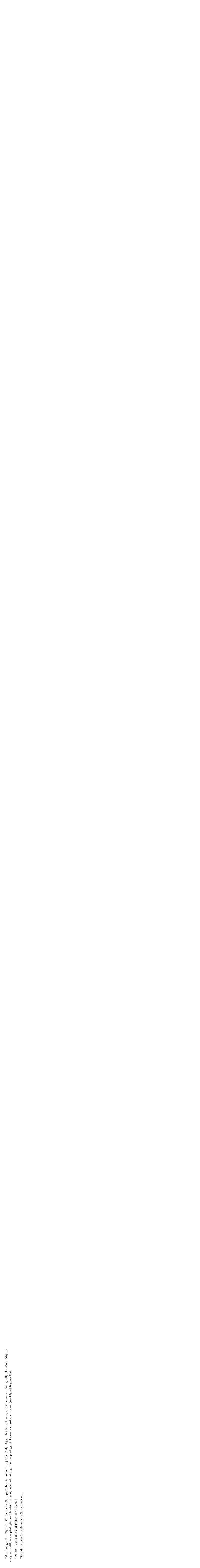}
\end{figure}

\clearpage

\begin{deluxetable}{cccccccc}
\tablewidth{0pt}
\tabletypesize{\scriptsize}
\tablecaption{Color-Magnitude Relation Fit Parameters\tablenotemark{a}\label{t_fitResults}}
\tablehead{\colhead{Color} & \colhead{Mag. Limit\tablenotemark{b}} & \colhead{$N_g$\tablenotemark{c}} & \colhead{Slope} & \colhead{Intercept} & \colhead{Offset\tablenotemark{d}} & \colhead{$\sigma_{\rm obs}$\tablenotemark{e}} & \colhead{$\sigma_{\rm int}$\tablenotemark{f}}}
\startdata
\sidehead{Morphologically selected sample (E+S0):}
$z_{850}-J$ & 22.5 & 19 & $-0.049\pm0.062$ & $1.335\pm0.046$ & $-0.005\pm0.037$ & $0.130\pm0.047$ & $0.123\pm0.049$\\
$z_{850}-K_s$ & 21.5 & 16 & $-0.221\pm0.057$ & $2.012\pm0.091$ & $-0.007\pm0.052$ & $0.177\pm0.051$ & $0.173\pm0.052$\\
\sidehead{Photometrically selected sample (all galaxies\tablenotemark{g}):}
$z_{850}-J$ & 24.0 & 34 & $-0.112\pm0.026$ & $1.282\pm0.016$ & $+0.014\pm0.025$ & $0.131\pm0.028$ & $0.118\pm0.034$\\
$z_{850}-K_s$ & 23.0 & 36 & $-0.299\pm0.021$ & $1.914\pm0.028$ & $-0.003\pm0.042$ & $0.241\pm0.035$ & $0.237\pm0.037$\\
\enddata
\tablenotetext{a}{Within 0.5 Mpc radius of the cluster X-ray position}
\tablenotetext{b}{In the redder passband of the color}
\tablenotetext{c}{Number of galaxies in the subsample}
\tablenotetext{d}{Biweight location estimate of color offset from the fitted relation}
\tablenotetext{e}{Biweight scale estimate of scatter about the fitted relation}
\tablenotetext{f}{Biweight scale estimate of scatter about the fitted relation, corrected for broadening by color errors}
\tablenotetext{g}{Only objects with $0.8 < z_{850}-J < 1.6$, $1.5 < z_{850}-K_s < 3.0$ (as appropriate) were included in the fit}
\end{deluxetable}

\begin{deluxetable}{cccccc}
\tablewidth{0pt}
\tabletypesize{\scriptsize}
\tablecaption{Inferred Stellar Population Ages\label{t_ages}}
\tablehead{\colhead{Color} & \colhead{Model\tablenotemark{a}} & \colhead{$t_f$\tablenotemark{b} (Gyr)} & \colhead{$z_f$\tablenotemark{c}} & \colhead{$\overline{t_L}$\tablenotemark{d} (Gyr)} & \colhead{$\overline{z_f}$\tablenotemark{e}}}
\startdata
\sidehead{Morphologically selected sample (E+S0):}
$z_{850}-J$   & BC03 & $> 1.3 \pm 0.8$ & $>2.1^{+0.7}_{-0.5}$ & $2.8 \pm 0.4$ & $4.0^{+1.1}_{-0.7}$\\
$z_{850}-J$   & M05  & $> 0.7 \pm 1.6$ & $>1.8^{+1.4}_{-0.6}$ & $2.5 \pm 0.8$ & $3.5^{+2.3}_{-1.0}$\\
$z_{850}-K_s$ & BC03 & $> 1.6 \pm 0.3$ & $>2.4^{+0.3}_{-0.3}$ & $3.0 \pm 0.2$ & $4.5^{+0.5}_{-0.4}$\\
$z_{850}-K_s$ & M05  & $> 0.5 \pm 1.2$ & $>1.7^{+0.8}_{-0.5}$ & $2.4 \pm 0.6$ & $3.3^{+1.3}_{-0.7}$\\
\sidehead{Photometrically selected sample (all galaxies\tablenotemark{f}):}
$z_{850}-J$   & BC03 & $> 1.3 \pm 0.5$ & $>2.2^{+0.4}_{-0.3}$ & $2.8 \pm 0.2$ & $4.1^{+0.7}_{-0.5}$\\
$z_{850}-J$   & M05  & $> 0.8 \pm 0.9$ & $>1.8^{+0.6}_{-0.4}$ & $2.5 \pm 0.4$ & $3.6^{+1.0}_{-0.6}$\\ 
$z_{850}-K_s$ & BC03 & $> 1.3 \pm 0.2$ & $>2.1^{+0.2}_{-0.1}$ & $2.8 \pm 0.1$ & $4.0^{+0.2}_{-0.2}$\\
$z_{850}-K_s$ & M05  & \nodata         & \nodata              & \nodata       & \nodata            \\
\enddata
\tablecomments{Entries in this table correspond to samples defined in Table~\ref{t_fitResults}. The measured internal scatter from the $z_{850}-K_s$ CMR of the photometrically selected sample is out of the range of the M05 model, and so is undefined.}
\tablenotetext{a}{Stellar population model: BC03 = \citet{BruzualCharlot_2003}, M05 = \citet{Maraston_2005}. See \textsection~\ref{s_discussionScatter} for details.}
\tablenotetext{b}{Minimum stellar population age at the redshift of J$2215.9-1738$.}
\tablenotetext{c}{Minimum formation redshift of the stellar population.}
\tablenotetext{d}{Mean luminosity weighted stellar population age at the redshift of J$2215.9-1738$.}
\tablenotetext{e}{Mean formation redshift corresponding to mean luminosity weighted stellar population age.}
\tablenotetext{f}{Only objects with $0.8 < z_{850}-J < 1.6$, $1.5 < z_{850}-K_s < 3.0$ (as appropriate) included.}

\end{deluxetable}


\end{document}